\documentclass[aps]{revtex4}
\usepackage{amsmath}
\usepackage{amssymb}
\usepackage{epsfig}
\usepackage{amscd}
\usepackage{graphicx,xspace}
\usepackage{float}
\usepackage[normalem]{ulem}

\begin{document}
%\begin{spacing}{1.5}
\title{SLE in self-dual critical $Z(N)$ spin systems: CFT predictions.} 
\author{Raoul
Santachiara} \email{raoul.santachiara@lpt.ens.fr} 
\affiliation{CNRS-Laboratoire de Physique Th\'eorique,
Ecole Normale
Sup{\'{e}}rieure\\  24 rue Lhomond, 75231 Paris Cedex 05, France}

\begin{abstract}
  The Schramm-Loewner evolution (SLE) describes the continuum limit of domain
  walls at phase transitions in two dimensional statistical systems. We
  consider here the SLE in  $Z(N)$ spin models at  their self-dual critical
  point. For $N=2$ and $N=3$ these models correspond to the Ising and
  three-state Potts model. For $N\geq 4$ the critical self-dual $Z(N)$ spin
  models are described in the continuum limit by non-minimal conformal field
  theories with central charge $c\geq 1$.  By studying the representations of the
  corresponding chiral algebra, we show that two particular operators satisfy
  a two level null vector condition which, for $N\geq 4$, presents an
  additional term coming from the extra symmetry currents action. 
 For $N=2,3$ these operators correspond to the boundary conditions changing
  operators associated to the SLE$_{16/3}$ (Ising model) and to the SLE$_{24/5}$ and
  SLE$_{10/3}$ (three-state Potts model). We suggest a definition of the
interfaces within the $Z(N)$ lattice models. The scaling limit of these interfaces is expected to be described at the self-dual critical point and  for $N\geq 4$ by the
SLE$_{4 (N+1)/(N+2)}$ and  SLE$_{4 (N+2)/(N+1)}$ processes.
\end{abstract}
\maketitle

\section{Introduction.}
The Schramm-Loewner evolutions (SLEs) are random growth processes that
generate conformally invariant curves in two dimensions (2D). SLEs yield
probability measures on the continuum limit of non-crossing interfaces in 2D statistical lattice models at criticality with conformal
invariance. SLEs has been proved successful to a more complete, and in some
case mathematically rigorous, description of fractal curves in critical
percolation \cite{Smirnov}, loop erased walk \cite{Lawler}, level lines of
height models \cite{Schramm} and domain boundaries at phase transitions (see e.g.
\cite{Bernard_review} and references therein).  The SLEs consider directly the
geometrical characterization of non-local objects in 2D critical systems and
complement the powerful tools provided by the conformal field theories (CFTs)
techniques.

On the other hand, CFTs focus on correlation functions of local operators
which are the scaling limit of lattice variables.  A first family of CFTs,
the minimal CFTs, calculates these correlation functions by studying the
infinite constraints imposed by the conformal invariance in 2D systems.
The Hilbert space of these theories is constructed from the highest weight
representations of the Virasoro algebra formed by  the generators of the
conformal symmetry.

The relation between SLEs and CFTs has been worked out in
\cite{Bernard_connection1,Bernard_connection2,Bernard_connection3}.  In this respect, an
important role is played by the boundary conditions changing operators (b.c.c.)  
which generate the boundary conditions from which the curve grows \cite{Cardy_bcft1,Cardy_bcft2}.
The key property of these operators is to satisfy particular relations under the action of the
conformal symmetry generators which lead to second order
differential equations for their correlation functions.  The link 
between the SLEs and CFTs is derived by comparing these second order equations
to the Fokker-Plank equations coming from the Brownian motion driving the SLE.

The SLEs/CFTs connection is well established in the case of minimal CFTs,
which have a central charge $c\leq 1 $. However, many critical models in
condensed matter and statistical physics posses, in addition to the
conformal invariance, symmetries in some internal degree of freedom, such as
the $SU(2)$ spin-rotational symmetry in critical quantum spin chains \cite{Affleck} or
the $Z_N$ symmetry in spin lattice models \cite{Zamo1,Zama_lat2}. In the continuum limit, these
additional symmetries are enhanced by the presence of chiral currents which
form, together with the energy-momentum tensor, more structured algebras which
present the Virasoro one as a sub-algebra.  The space of local fields occurring
in a CFT with additional symmetry, the non-minimal CFT,  corresponds to the representations of the
associated chiral algebra. These theories have in general a central charge
$c\geq 1$.

The  connection between SLEs and non-minimal CFTs  has been first addressed in \cite{Rasmussen1,Rasmussen2}, where the relation between stochastic evolutions  and superconformal field theory was investigated. More recently,  the connection between SLE and Wess-Zumino-Witten models, i.e. CFTs with additional Lie-group symmetries,  has been studied  by very different approaches    \cite{Rasmussen3, Ludwig}. In particular,  it has been shown   in \cite{Ludwig}  that a
consistent SLE approach is possible provided that an additional stochastic
motion in the internal symmetry group space in introduced. For an $SU(2)$ symmetry, for
instance, the SLE describes a process which carries a fluctuating additional spin
$1/2$ degree of freedom.

Inspired by these results, we discuss in this paper a possible SLE approach to
CFTs with additional $Z_N$ symmetry. The conserved currents associated to the
extra symmetry, called parafermions, have fractional spin.  We consider the
first of such parafermionic theories with central charge $c=2(N-1)/(N+2)$
\cite{Zamo1}.  The $Z(2)$ and $Z(3)$ models coincide with some particular
minimal CFT where the domain wall boundary conditions and the associated
b.c.c.  operator are known. We exploit these pieces of information to
identify the possible b.c.c. operators for general $Z_N$ symmetry.
We show that, under the action of the parafermionic conserved currents, these
operators satisfy algebraic relations similar to the ones found in the case of
minimal models. From these CFT predictions and in analogy with the approach
proposed in \cite{Ludwig}, we introduce an additional stochastic motion driven
by the action of the parafermionic currents.

The strong motivation to study these theories lies on the fact that
they describe the continuum limit of critical lattice models, the $N$-states
generalization of the Ising model which reduce to the Ising and three
state Potts model for $N=2$ and $N=3$. 
   This allows for  the identification of the  SLE interfaces on the lattice and opens the possibility of  a direct numerical verification.

The paper is organized as follows. We review the connection between SLE and
minimal CFTs in Section \ref{sle_review}. Section \ref{slelie} provides the
extension to CFTs with extra Lie symmetries. In Section \ref{sle/para}, after
an introduction to the operator content of the parafermionic theories, we
define a stochastic motion in the $Z_N$ internal space. Further, by studying
the parafermionic algebra representations, we derive some algebraic relations
related to the b.c.c. operators. In Section \ref{lattice
  model} we shall discuss the definition of interfaces on the lattice.
We conclude the paper in Section \ref{conclusion}.

\section{Stochastic Loewner evolutions and conformal field theories.}
\label{sle_review}
\subsection{The SLE$_{\kappa}$/Minimal models connection.}

\subsubsection{Chordal SLE: definition.}
Throughout this paper, we consider chordal SLE which describes random curves
joining two boundary points of a connected planar domain.  For a comprehensive
and detailed introduction to SLE, see e.g.
\cite{Walter,Cardy_review,Bernard_review}. The definition of SLE is most
conveniently given in the upper half complex plane $\mathbb{H}$: it describes
a fluctuating self-avoiding curve $\gamma_{t}$ which emanates from the origin
($z=0$) and progresses in a properly chosen time t. If $\gamma_{t}$ is a
simple curve, this evolution is defined via the conformal map $g_t(z)$ from
the domain $\mathbb{H}_t=\mathbb{H}/\gamma_{]0,t]}$, i.e. the upper half plane
from which the curve is removed, to $\mathbb{H}$. In the more general case of
non-simple curves, the function $g_t(z)$ produce conformal maps from
$\mathbb{H}_t=\mathbb{H}/K_t$ to $\mathbb{H}$ where $K_t$ is the SLE hull at
time $t$.  The SLE map $g_t(z)$ is uniquely determined by fixing three real
parameters.  A conventional normalization (the so called hydrodynamic
normalization) is such that $g_t(z)=z+2t/z+\cdots$ near $z=\infty$. With this
choice the time $t$ corresponds to the upper half-plane capacity $c_{K_t}$ of
the hull $K_t$, $c_{K_t}=2t$. The function $c_{K_t}$ is a positive quantity
and satisfies the additive law $c_{K_{t}\cup
  g_{t}^{-1}(K_{t'})}=c_{K_{t}}+c_{K_{t'}}$. The SLE map $g_t(z)$ is a
solution of the Loewner equation:
\begin{equation}
  \frac{d}{d t} g_t (z)=\frac{2}{g_t(z)-\xi_t} \quad g_{t=0}(z)=z,
\label{Sle_definition}
\end{equation}
where $\xi_t$ is a real valued process, $\xi_t \in \mathbb{R}$, which drives the evolution
of the curve. For a system which satisfies the Markovian and conformal
invariance properties, together with the left-right symmetry, the process
$\xi_t$ is shown \cite{Schramm_2} to be proportional to a Brownian motion: {\bf E}$[\xi_t]=0$
and {\bf E}$[\xi_t \xi_s]= \kappa\,\mbox{min}(s,t)$. In the following, we use the
symbol {\bf E}[...] for the stochastic average over the Brownian
motion.

\subsubsection{Martingales in SLE processes.}
In
\cite{Bernard_connection1,Bernard_connection2,Bernard_connection3,Bernard_connection4}
it was understood how to associate CFT states with the growing curves of the
Loewner process.  To reveal the SLE/CFT connection, the first step is to
define stochastic process such that conditioned correlation function are
martingales for this process.  The main idea is to arrange the statistical sum
by first summing over all the configurations $\mathcal{C}_{\gamma_{t}}$
presenting a trace $\gamma_{t}$ of the interface with fixed shape and capacity 
$t/2$ and then summing aver all the possible shapes of the trace. The first sum
defines a conditioned correlation function while the second sum can be seen as
the stochastic mean of this correlation function. To make it more concrete,
consider an observable $\mathcal{O}$ of a lattice model defined on the domain
$\mathbb{H}$. The statistical sum $\prec O\succ_{\mathbb{H}}$ can be written
as:
\begin{equation}
  \prec \mathcal{O} \succ_{\mathbb{H}}=
\mbox{{\bf E}}[\prec \mathcal{O} \succ _{\gamma_{t}}]
=\sum_{\gamma_{t}}P[\mathcal{C}_{\gamma_{t}}]\prec
  \mathcal{O} \succ _{\gamma_{t}},
\label{martingaletrick}
\end{equation}
where $\prec 0 \succ_{\gamma_{t}}$ is the statistical average conditioned to
the presence of the trace $\gamma_{t}$ and $P[\mathcal{C}_{\gamma_{t}}]$ is
the probability of its occurrence.  The correlator $\prec
\mathcal{O}\succ_{\mathbb{H}}$ does not depend on the choice of $t$: the
stochastic mean of the correlator $\prec 0 \succ_{\gamma_{t}}$ is thus time
independent and it is a martingale.  

\subsubsection{Martingales and conformal correlators.}
The relation (\ref{martingaletrick}) becomes extremely useful at the critical
point where the model is expected to be described in the continuum limit by a
conformal field theory. In this case, the
operator $\mathcal{O}$ corresponds generally to a product of, say, $l$ primary
fields $\phi_i(z_i)$, $\mathcal{O}(\{z_i\})=\prod_{i=1}^{l} \phi_i(z_i)$ at
positions $z_i$. The image $^{f}\phi$ of $\phi$ under a conformal transformation
$f(z)$ is
\begin{equation}
^{f}\phi_i=[\partial_z f(z)]^{\Delta_{i}}\phi_i(f(z_i)),
\label{primary_cond}
\end{equation}
where $\Delta_{i}$ is the conformal dimension of $\phi_{i}$.  The statistical
expectation values can then be expressed in terms of CFT correlation functions
$\mathcal{F}(\{z_i \})_{\mathbb{H}_{t}}$:
\begin{equation}
  \prec \mathcal{O} \succ_{\gamma_{t}}
  \rightarrow \mathcal{F}(\{z_i\})_{\mathbb{H}_{t}}=\frac{<\mathcal{O}(\{z_i\}) \psi
    (\infty)\psi(z_t)>_{\mathbb{H}_{t}}}{< \psi
    (\infty)\psi(z_t) >_{\mathbb{H}_{t}}}
\label{martingale_cft}
\end{equation}
where $<\dots>_{\mathbb{H}_{t}}$ indicates the conformal correlation function
computed in the domain $\mathbb{H}_{t}=\mathbb{H}/\gamma_t$, i.e. the upper
half plane with the trace $\gamma_t$ removed.  The fields $\psi(z_t)$ and
$\psi(\infty)$, inserted respectively at the tip $z_t$ of $\gamma_{t}$ and at
the infinity, are the b.c.c. operators implementing the boundary
conditions at which the interface anchor.

Using the conformal invariance, the correlation functions
$\mathcal{F}(\{z_i\})_{\mathbb{H}_{t}}$ in the domain wall ${\mathbb{H}_{t}}$
can be expressed as the correlation functions in the upper-half plane
${\mathbb{H}}$:
\begin{equation}
 \mathcal{F}(\{z_i\})_{\mathbb{H}_{t}}=\frac{<^{g_t}\mathcal{O} (\{z_i \})\, \psi
    (\infty)\,\psi(\xi_t)>_{\mathbb{H}}}{< \psi
    (\infty)\,\psi(\xi_t) >_{\mathbb{H}}}. 
\label{martingale_cft_up}
\end{equation}
Note that the Jacobians coming from the conformal transformation on the $\psi$
fields cancel between the numerator and the denominator in the above expression.

We are now in the position to understand the SLE/CFT connection.  Under the
SLE, the trace $\gamma_t$ evolves and the iterated sequence of infinitesimal
conformal mappings $g_t(z)$ satisfying Eq.(\ref{Sle_definition}) leads to a
Langevin dynamics for the conformal correlator
$\mathcal{F}(\{z_i\})_{\mathbb{H}_{t}}$.  Imagine evolving the trace for a
time $t$ and then, for an infinitesimal time $dt$. Using
Eq.(\ref{Sle_definition}), the variation $d (^{g_t}\phi_i(z_i))=^{g_{t+d
    t}}\phi_i(z_i)\, -\, ^{g_{t}}\phi(z_i)$ is given by: 
\begin{equation}
  d (^{g_t}\phi_i(z_i))=\frac{1}{2\pi i}\oint_{z_i} d g_t(w) T(w)\, ^{g_t}\phi_i(z_i)=
2 dt \left(-\frac{\Delta_i}{(g_t(z_i)-\xi_t)^2}+\frac{\partial_{g_t(z_i)}}{g_t(z_i)-\xi_t}
\right)\, ^{g_t}\phi_i(z_i)
\label{spect_dynamic}
\end{equation}
where $T(z)$ is the energy-momentum tensor. The variation for the $\psi$'s are
given by the Ito differential:
\begin{equation}
d (\psi(\xi_t)) =  \partial_{\xi_t}\psi(\xi_t) d \xi_t +\frac{\kappa}{2}
\partial^2_{\xi_t} \psi (\xi_t) dt.  
\label{psi_dynamic}
\end{equation} 
Using the Eqs.(\ref{psi_dynamic})-(\ref{spect_dynamic}) in
Eq.(\ref{martingale_cft_up}) and averaging over all the realization of
$\gamma_t$, one obtains the diffusion equation:
\begin{equation}
  \partial_{t} \mbox{{\bf E}}[\mathcal{F}(\{z_i\})_{\mathbb{H}_{t}}]=\left(2\sum_i\left[-\frac{\Delta_i}{(g_t(z_i)-\xi_t)^2}+\frac{\partial_{g_t(z_i)}}{g_t(z_i)-\xi_t}\right]+\frac{\kappa}{2}\partial^2_{\xi_t} 
 \right)\mbox{{\bf E}}[\mathcal{F}(\{z_i\})_{\mathbb{H}_{t}}].
\label{diff_equation}
\end{equation}
As previously said, the correlation function
$\mathcal{F}(\{z_i\})_{\mathbb{H}_{t}}$ is a martingale of the SLE process and
satisfies therefore the following differential equation
\begin{equation}
\left(2\sum_i\left[-\frac{\Delta_i}{(g_t(z_i)-\xi_t)^2}+\frac{\partial_{g_t(z_i)}}{g_t(z_i)-\xi_t}\right]+\frac{\kappa}{2}\partial^2_{\xi_t} 
 \right)\mathcal{F}(\{z_i\})_{\mathbb{H}_{t}}=0.
\label{diff_eq_F}
\end{equation}

\subsubsection{Operator formalism and minimal models.}
In order to show the consequences of the above relation, it is convenient to
introduce the operators $L_n$ which are the modes of the energy-momentum
tensor $T(z)$:
\begin{equation}
L_n \phi(z)=\frac{1}{2\pi i} \oint_{z} d w\, w^{n+1} T(w) \phi(z).
\end{equation}
The $L_n$ are the generators of the conformal transformation in the CFT
Hilbert space. A primary operator $\phi$, satisfying Eq. (\ref{primary_cond}),
is annihilated by the positive modes of 
$L_n$,  $L_n \phi=0$ for $n>0$,  and the eigenvalue of the zero mode is its
conformal weight, $L_0 \phi =\Delta \phi$.
The $L_{n}$ satisfy the Virasoro
algebra
\begin{equation}
[L_n,L_m]=(n-m)L_{n+m} + \frac{c}{12}n^2(n-1)\delta_{n+m}. 
\label{Virasoro}
\end{equation}
All local fields of the theory can be obtained by applying the $L_{-n}$,
$n\geq 1$, to a primary fields $\phi$. Each set of states of the form
$L_{k_1}\cdots L_{k_m} \phi$ forms a conformal family $[\phi]$ and corresponds
to the representation of the highest weight of the Virasoro algebra, also
called Verma module, the primary field $\phi$ corresponding to the highest
vector.  The descendant state $L_{-k_1}\cdots L_{-k_m} \phi$, $\sum k_m= n$, is
said to be the $n-$th level of the module of $\phi$ and its conformal weight is
$\Delta+n$.

For $c\leq 1$, the unitary representations of the Virasoro algebra has central
charge \cite{Friedan}:
\begin{equation}
c=1-\frac{6}{m(m+1)}\quad m=2,3,\cdots
\label{c_minimal}
\end{equation}
and correspond to the simplest family of conformal theories, called minimal
models $M_m$. The minimal model $M_m$ is characterized by a finite number of
Verma modules $[\phi_{r,s}]$ with conformal weight:
\begin{equation}
\Delta_{r,s}=\frac{[(m+1)r-m s]^2 -1}{4 m(m+1)}, \quad r=1,2,\cdots m-1, \quad
s=1,2,\cdots r.
\label{kac_minimal}
\end{equation}

\subsubsection{Null vectors in CFT and SLE. }
In CFT, the differential equations satisfied by the correlation function can
be derived from the structure of the representation modules and, in
particular, from  the relations between the descendant states. 
One can show that $\mathcal{F}(\{z_i\})_{\mathbb{H}_{t}}$ is a solution of 
(\ref{diff_eq_F} ) 
if the b.c.c. operator $\psi$ obeys the
condition:
\begin{equation}
(L_{-2} -\frac{\kappa}{4}L_{-1}^2) \psi=0,
\label{sle_nul_vect}
\end{equation}
The equation Eq.(\ref{sle_nul_vect}) means that the descendant states
$L_{-1}^2 \psi$ and $L_{-2} \psi$ are linearly dependent. The operator $\psi$
transforms then as a degenerate representation of the Virasoro algebra which
has a null state at level 2.  The operator $\psi$ can thus be identified with
the operators $\phi_{1,2}$ or $\phi_{2,1}$ of the minimal models $M_m$ which
are shown to satisfy:
\begin{eqnarray}
(L_{-2} - \frac{3}{2(2 \Delta_{1,2}+1)} L_{-1}^2) \phi_{1,2}&=&0,
\label{vir_nul_vect_1} \\
(L_{-2} - \frac{3}{2(2 \Delta_{2,1}+1)} L_{-1}^2)\phi_{2,1}&=&0.
\label{vir_nul_vect_2}
\end{eqnarray}
Comparing Eq.(\ref{vir_nul_vect_1}) and Eq.(\ref{vir_nul_vect_2}) to Eq.(\ref{sle_nul_vect}) and using
Eq.(\ref{c_minimal}) and Eq.(\ref{kac_minimal}), one can finally make explicit
that the SLE/CFT connection: an SLE process with parameter $\kappa$ describes
interfaces in CFT with central charge:
\begin{equation}
c_{\kappa}=\frac{(6-\kappa)(3\kappa-8)}{2 \kappa}
\label{sle_cft_minimal},
\end{equation}
and  the boundary conformal operator $\psi$ has scaling dimension:
\begin{equation}
\Delta_{\kappa}=\frac{(6-\kappa)}{2 \kappa}
\label{sle_cft_bcc},
\end{equation}
This derivation shows that martingales for SLE
processes are closely related with the existence of null vectors in the the
appropriate Verma module.

\section{SLEs for CFTs with extra Lie group symmetries.}
\label{slelie}

As mentioned above, the operator content of the minimal models $M_m$, which
have central charge $c\leq 1$, corresponds to the space of representations of
the Virasoro algebra (\ref{Virasoro}). In this case, the conformal symmetry
determines completely the spectrum of the theory and the differential
equations satisfied by the correlation functions.

The applications of the SLEs defined in the previous section are limited to
CFTs with central charge $c\leq 1$. This can be directly understood from the
fact that, for a CFT with extended symmetry, not all the local fields can be
obtained by the application of the $L_n$. In particular, one expects at each
level additional states and the  linear relations
 of the type (\ref{vir_nul_vect_1})-(\ref{vir_nul_vect_2}) will include additional terms.

\subsection{SLEs and CFTs with Lie group symmetries.}
The $SU(2)$ WZW models are among the most important CFTs with additional
symmetries.  These CFTs posses an internal $SU(2)$ continuous symmetry which,
together with the conformal one, is realized by the set of chiral currents
$J^{a}(z)$, $a=+,-,0$, with scaling dimension $\Delta_{J}=1$.  The current
algebra, derived from the current-current operator product expansion (OPE),
takes the form:
\begin{equation}
[J^{a}_{n},J^{b}_{m}]=f^{a,b}_{c} J^{c}_{n+m}+N\, n \,\delta_{n+m},
\label{kac_moody}
\end{equation}
where $J^{a}_n$ are the current modes, $J^{a}_{n}=1/(2\pi i)\oint_{z} d w
\,w^n \,J^{a}(w)$, the $f^{a,b}_{c}$ are the structure constants of the
$su(2)$ Lie algebra and $N$ is the level of the algebra. The representations
of the Kac-Moody algebra , leads to a family of CFTs, denoted
$SU(2)_{N}$, with central charge $c_N=3N/(N+2)$. 
The local fields $\phi_{j}(z)$ of
the theory transform under the action of the chiral currents in the
representation space of dimension $j(j+1)$ of $SU(2)$. Analogously to the
Virasoro case, a WZW primary field is defined by:
\begin{equation}
J^{a}_{n}\phi_{j}(z)=0 \quad \mbox{for}\,\, n>0; \quad J^{a}_{0}\phi_{j}(z)=t^{a}_{j}\phi_{j}(z),
\end{equation}
where the $t^{a}_{j}$ are the $SU(2)$ generator matrices in the $j$
representation. 
The conformal weight $\Delta_{j}$ of a primary field  $\phi_{j}(z)$ is 
\begin{equation}
L_0 \phi_{j}(z)=\Delta_{j}\phi_{j}(z)= \frac{j(j+1)}{N+2}\phi_{j}(z) .
\label{dim_wzw}
\end{equation}
Under an infinitesimal conformal $d g(z)$ transformation together with an
infinitesimal $SU(2)$ gauge transformation $d \theta^{a}(z)$, the primary
fields $\phi_{j}(z)$ changes as:
\begin{equation}
d (\phi_{j})(z)=\frac{1}{2\pi i}\oint_{z} d\,w\,d g (w) T(w)\, \phi_{j}(z)+ \frac{1}{2\pi i}\oint_{z} d\,w\,d \theta^{a} (w) J^{a}(w)\, \phi_{j}(z)
\label{wzw_varia}
\end{equation}
 
In \cite{Ludwig}, the SLE approach has been extended to these theories. It
proposes  to describe this
model by a composition of two independent Brownian motions,
one in the physical space, defined as in Eq.(\ref{Sle_definition}), and one in the
internal $SU(2)$ spin degrees of freedom: 
\begin{equation}
d \theta^{a}_t(z)=\frac{d \theta^{a}_t}{g_t(z)-\xi_t}, \quad \mbox{{\bf
    E}}[\theta^{a}_t\theta^{b}_s]=\tau\,\delta^{a,b}\,\mbox{min}(s,t), 
\label{tau_sle}
\end{equation}
where the variance $\tau$ is an independent parameter of the SLE evolution.
The idea is then to attach to trace $\gamma_t$ a spin $1/2$ degrees of
freedom. The interface then undergoes both a standard SLE evolution in the
physical space and a stochastic $SU(2)$ rotation.

The derivation of the SLE/CFT connection is strictly analogous to the one
shown before. The difference here is that the b.c.c. operator
$\psi(\xi_t)$ in Eq.(\ref{martingale_cft_up}) carries a spin $1/2$,
$\psi\equiv\psi_{1/2}(\xi_t)$ which transforms under the representation
$\phi_{1/2}$. Using Eq.(\ref{Sle_definition}) and Eq. (\ref{tau_sle}) in
Eq.(\ref{wzw_varia}), the Ito formula for $\psi_{1/2}(\xi_t)$, written in terms of
the modes $L_n$ and $J^a_n$, is:
\begin{equation}
d( \psi_{1/2}(\xi_t))= d \xi_t L_{-1}\psi_{1/2}(\xi_t)  + d \theta^{a} J^{a}_{-1} \psi_{1/2}(\xi_t)
+ dt\left( \frac{\kappa}{2}
L_{-1}^2 +\frac{\tau}{2} J^{a}_{-1}J^{a}_{-1}\right)\psi_{1/2}(\xi_t).
\label{wzw_psi_dinamic}
\end{equation}
The operator $\mathcal{O}$ is now a product of WZW primaries and remain fixed
while the trace evolves. This means one has only to consider their variation
(\ref{spect_dynamic}) under the conformal transformation $g_t$ which maps the
domain $\mathbb{H}_t$ in the domain $\mathbb{H}_t$.  The martingale condition
(\ref{sle_nul_vect}) takes the form:
\begin{equation}
\left(L_{-2} -\frac{\kappa}{4}L_{-1}^2-\frac{\tau}{4}J^{a}_{-1}J^{a}_{-1}\right) \psi_{1/2}=0.
\label{sle_nu_vect_wzw}
\end{equation}
The correlation functions in WZW theories satisfy first order differential
equations, called Knizhnik-Zamolodchikov equations, which are derived from a
relation between descendant states at the first level:
\begin{equation}
\left(L_{-1}+\frac{1}{N+2}J^{a}_{-1}t^{a}\right)\phi^{m}_{j}=0.
\label{KZ}
\end{equation}
Applying the operator $L_{+1}$ to Eq.(\ref{sle_nu_vect_wzw}) and using the
commutation relation (\ref{kac_moody}) together with $[L_{n},J^{a}_m]=-m
J^{a}_{n+m}$, one fixes a first relation between $\kappa$, $\tau$ and $N$,
$\tau/(6-\kappa(2\Delta_{1/2}+1))=2/(N+2)$. It was shown \cite{Ludwig} that a second
condition, $\kappa+\tau=4$, can be obtained by demanding the
one-point function of the current to exhibit a simple pole at the tip of the
trace $z_t$. The final result is:
\begin{equation}
\kappa=4\frac{N+2}{N+3}\quad \tau=\frac{4}{N+3}.
\end{equation}

\section{SLEs in $Z(N)$ parafermionic theories.}
\label{sle/para}
We consider the CFTs with extended $Z_N$ symmetries, the so calles $Z(N)$
parafermionic theories. In particular, we focus on the $Z(N)$ parafermionic
theories with central charge $c=2(N-1)/(N+2)$ ($c>1$ for $N>4$). These
theories describe the continuum limit of the $N$-states spin models
interacting via a $Z_N$
invariant nearest-neighbor coupling, at their
self-dual critical points (see section \ref{lattice model}).  For $N=2$ and
$N=3$ one finds the well known Ising and three-state Potts model. The
corresponding CFTs has central charge $c=1/2$ and $c=4/5$ respectively and
coincide with the minimal models $M_3$ and $M_5$: the operator content of
these theories can also be determined by studying the representations of the
Virasoro algebra.  In this sense, the $Z_2$ and $Z_3$ symmetry of these models
are trivially realized.  The CFTs describing the critical point of the Ising
and the three-state Potts are then expected to be described by the SLE in
Eq.(\ref{Sle_definition}). The b.c.c.  conformal operators 
implementing the SLE interface boundary conditions have been identified.  We
will show that these results admit a natural extension to general $N$.

\subsection{Parafermionic current algebra.}
We briefly review the $Z(N)$ parafermionic theories with central
charge $c=2(N-1)/(N+2)$.  These theories were introduced and constructed in
\cite{Zamo1,Zamo2}. Here we shall enounce the main results in a slightly
different manner.
The general arguments and the notations used here are strictly analogous to the
ones discussed  in a series of papers \cite{Raoul1,Raoul2,Raoul3,Raoul4}  where a second series of
parafermionic theories was studied. 

Extra $Z_N$ group symmetries in two-dimensional conformal field theories are
generally realized by a set of holomorphic currents $\Psi^{k}(z)$
($k=1,\cdots,N-1$), satisfying the following operator product expansion:
\begin{eqnarray}
\Psi^{k}(z)\Psi^{k'}(z') &=& 
\frac{\lambda^{k,k'}_{k+k'}}{(z-z')^{\Delta_{k}^{\Psi}
 +\Delta_{k'}^{\Psi}-\Delta_{k+k'}^{\Psi}}} \label{opepara1} \\
 &\times& \left \{\Psi^{k+k'}(z')+0(z-z')\right \},\quad k+k'\neq 0 \nonumber \\
 \Psi^{k}(z)\Psi^{-k}(z') &=& \frac{1}{(z-z')^{2\Delta_{k}^{\Psi}}}
 \left \{1+(z-z')^{2}
 \frac{2\Delta_{k}^{\Psi}}{c}T(z')+\ldots \right \}
 \label{opepara2}
\end{eqnarray}
where $\Delta_{k}^{\Psi}$ is the conformal dimension of the parafermions
$\Psi_{k}(z)$ and $\lambda_{k+k'}^{k,k'}$ are the structure constants of the
algebra. We shall be interested in parafermionic theories in which the
dimensions $\Delta_{k}^{\Psi}$ of the parafermions $\{\Psi^{k}\}$ take the minimal possible values
admitted by the associativity constraint:
\begin{equation}
 \Delta_{k}^{\Psi}=\Delta_{N-k}^{\Psi}=\frac{k(N-k)}{N} \quad k=0,1\cdots N-1.
 \label{chidim}
\end{equation}
Note that the above formula is not symmetric under the exchange $k\to -k$ :
the field $\Psi^{-k}$ in (\ref{opepara2}) is assumed to have
dimension $\Delta_{N-k}^{\Psi}$, in the sense that the indices $k$ referring
to the $Z_N$ charge are always defined modulo $N$. Thus,
\begin{equation}
 \Psi^{N-k}\equiv\Psi^{-k}\equiv(\Psi^{k})^{+},\quad \Delta_{N-k}^{\Psi}
 \equiv\Delta_{-k}^{\Psi}.
\end{equation}
The structure constants $\lambda_{k+k'}^{k,k'}$ and the central 
charge
$c$ (of the Virasoro algebra) are given by the expressions:
\begin{eqnarray}
  (\lambda^{k,k'}_{k+k'})^{2} &=&
  \frac{(k+k')!(N-k)!(N-k')!}{k!k'!(N-k-k')!N!} \label{struct} \\
  c &=& \frac{2(N-1)}{N+2}. \label{c}
\end{eqnarray}

\subsection{Representation space: the $Z_N$ sector.}

The CFTs we are considering describes the self-dual (Kramers-Wannier
invariance) critical point of $Z_N$-invariant lattice model \cite{Zamo1}. In addition to
the $Z_N$ symmetry, the theory possesses the dual $\tilde{Z}_N$ invariance,
which is enanched by the conservation of  the antiholomorphic fields $\overline{\Psi}(\overline{z})$. The
Hilbert space of a $Z_N\times \tilde{Z}_N$ invariant theory, splits into
subspaces characterized by the $Z_N \times \tilde{Z}_N$ charges $\{p,q\}$. A
field $\phi_{\{p,q\}}$ in one of these subspaces transform as
$\phi_{\{p,q\}}(z)\to \exp[2i\pi (p m+ q n)/N]\phi_{p,q}(z)$ under global
rotations of angles $2\pi m/N$ and $2\pi n/N $.  The lattice spin operator
$\sigma_k$ (see \ref{lattice model}) and its dual $\mu_k$ are described in the continuum limit
by fields of charge $\{k,0\}$ and $\{0,k\}$ respectively. The
(anti-)holomorphic currents $\Psi^{k}(z)$ ( $\overline{\Psi}(\overline{z})$),
which appear in the OPE of $\sigma_k \mu_k$ ($\sigma_k \mu_k^{+}$), have
charges $\{k,k\}$ ($\{k,-k\}$). Recently, the lattice holomorphic realization
of the parafermionic currents has been discussed in \cite{Riva}.

It is in general
convenient to express the charge ${k,k'}$ as $[q^{*},p^{*}]=\{k+k',k-k'\}$, where now
$q$ and $p$ are defined mod $2N$ and $q^{*}+p^{*}$ even. In this notation, the
currents $\Psi^{k}$ and $\overline{\Psi}(\overline{z})$ have charge $[2k,0]$
and $[0,2k]$ respectively. 

In the following we concentrate on the action of the holomorphic field
$\Psi^{k}(z)$, all the results being valid also for
$\overline{\Psi}(\overline{z})$. Thus, without losing any generality, one can
study the structure of the $Z_N$ representations which we denote with
$\Phi^{\pm q^{*}}$. Moreover, we find convenient to use a different convention
$q$ for the $Z_N$ charges, defined by $q^{*}= 2 q \,\mbox{mod} 2 N$. With this
choice, the product $\Psi^{k}(z)\Phi^{\pm q}$ is a field with
charge $k \pm q$.

For $N$ odd, we consider then the representation fields
\begin{equation}
\Phi^{\pm q}(z,\bar{z})\quad q=0,\pm 1,\cdots,\pm (N-1)/2 \quad(N \,\mbox{odd}).
\end{equation} 

For $N$ even, it turns out that the modules of the representation
corresponding to $\Phi^{\pm q}$ with $\lfloor N/4 \rfloor <q \leq \lfloor N/2
\rfloor$ are identical to those of $0 \leq q \leq \lfloor N/4 \rfloor$. In
order to recover the right number of $Z_N$ representations one has to consider
half-integer value of $q$:
\begin{equation}
\Phi^{\pm q}(z,\bar{z})\quad q=0,\pm \frac{1}{2},\pm 1,\cdots,\pm \lfloor N/4 \rfloor \quad(N \,\mbox{even}).
\end{equation} 
Naturally, the physical meaning of the the half-integer charge is recovered in
the usual notation $q^{*}$.

\subsubsection{Currents modes in $Z_N$ sector.}

The currents $\{\Psi^k\}$  can be decomposed into mode operators
$A^{k}_{\cdots+n}$, whose action is to change the $Z_N$ charge of the representation fields $\Phi^{q}$: 
\begin{eqnarray}
\Psi^{k}(z)\Phi^{q}(0)&=& \sum_{n}\frac{1}{(z)^{\Delta_{k}^{\Psi}-\delta_{k}^{k+q}+n}}
  A^{k}_{-\delta^{q}_{k+q}+n}\Phi^{q}(0) 
\label{eq3} \\
\delta^{q}_{k}&=&\frac{q^{2}-k^{2}}{N} \mbox{ mod }\, 1.
\label{level_structure}
\end{eqnarray}
The value $\delta^{q}_{k}$ is the first level in the module of
$\Phi^{\pm q}$ corresponding to the $Z_N$ charge $k$: it determines thus the
level structure of the modules induced by the $Z_N$ representation fields. 
The values of $\delta^{q}_{k}$ can be easily obtained by considering the
module of the identity whose descendant states are the chiral currents
$\Psi^{k}(z)$. The levels of these operators correspond to their conformal
dimensions $\Delta_{\pm k}^{\Psi}$. Taking into account that, owing to the Abelian
monodromy of the fields $\Psi^{k}(z)$ in the $Z_N$ sector, the level spacing
is equal to 1, one can readily obtain from Eq.(\ref{chidim})
$\delta^{0}_{k}=-k^2/N \,\mbox{mod}\,1$. The level structure of
a generic field $\Phi^{\pm q}(z)$, Eq.(\ref{level_structure}), is then
extracted from the module of the identity by inspecting its corresponding
submodule. Note also that $\delta^{q}_{k+q}=0$ for $k=-2 q$.  This results in the
presence of zero modes $A^{\pm q}_{0}$:
\begin{equation}
 A^{\mp 2q}_{0}\Phi^{\pm q}(0)=h_{q}\Phi^{\mp q}(0). \label{zeroeigen}
\end{equation}
which associate at each field $\Phi^{q}$, with conformal dimension
$\Delta_{q}^{\Phi}$,  the field $\Phi^{-q}$ with
opposite charge and with the same conformal dimension, $\Delta_{q}^{\Phi}=\Delta_{-q}^{\Phi}$. 
The eigenvalues $h_q$ defined in (\ref{zeroeigen}) characterize the representations of the
parafermionic algebra together with the conformal dimension the fields $\Phi^{q}$.
As usual, primary fields are defined by
$A^{k}_{-\delta^{q}_{k+q}+n}\Phi^{q} = 0$ for $n>0$.  
Each representation module is then characterized by the two primary fields $\Phi^{\pm q}$ which can be obtained
one from the other as in (\ref{zeroeigen}). 
An example of the module structure associated to the field $\Phi^{\pm 1}$ and of the action of the currents is given in
Fig.\ref{phi1modulo}  for $N=5$. Using the formula (\ref{level_structure}), we
have $\delta^{\pm 1}_{0}=1/5$ and $\delta^{\pm 1}_{\pm2}=4/5$.

Using the expansion (\ref{eq3}), the action of the modes in each sector can be given in terms of a
contour integral:
\begin{equation}
 A^{k}_{-\delta^{q}_{k+q}+n}\Phi^{q}(0) = \frac{1}{2\pi i}\oint_{C_{0}}
 {\rm d}z \,(z)^{\Delta_{k}^{\Psi}-\delta^{q}_{k+q}+n-1}\Psi^{k}(z)\Phi^{q}(0).
 \label{modes}
\end{equation}
In the $Z_N$ sector, one can simply consider the action of the currents
$\{\Psi^{\pm 1} \}$ because, via the Eq. (\ref{opepara1}), they completely
determine the full algebra of the other currents $\{\Psi^{\pm k}\}$,
$k=2,\cdots, \lfloor N/2 \rfloor$.  

The commutation relations of the mode operators can be
deduced from Eq.(\ref{modes}) by using standard techniques in the complex
plane. These relations are given in Appendix \ref{commutation_relation_zn}.

As shown in the Appendix \ref{commutation_relation_zn}, the representation
space contains $\lfloor N/2 \rfloor$+1 primary operators $\Phi^{q}$ with
$q=0,1,\cdots, \frac{N}{2}$ for $N$ odd and with $q=0,\pm \frac{1}{2},\pm
1,\cdots,\pm \lfloor N/4 \rfloor$ for $N$ even.  The conformal dimension of
these operators turn out to be:
\begin{equation}
\Delta_{q}^{\Phi}=\frac{q(N-2 q)}{N(N+2)} \quad \, q=0,
1,\cdots, \frac{N-1}{2} (N\,\mbox{odd}),\quad q=0, \frac{1}{2}, 1,\cdots,  \frac{N}{4}  \quad(N \,\mbox{even}) .
\label{kac_zn}
\end{equation}
As said above, to each primary $\Phi^{q}$ corresponds another primary field
$\Phi^{-q}$ of the same dimension.

In the case $N=2$ and $N=3$ (i.e. Ising and three state Potts model), one has
only two primaries of the parafermionic algebra. For $N=2$ one finds the
identity operator with $\Delta_{0}=0$ and the operator $\Phi^{\pm 1/2}$, with
dimension $\Delta_{\pm 1/2}=1/16$, corresponding to the Ising model spin
operator. Analogously, for $N=3$ one has, together with the identity operator,
the operator $\Phi^{\pm 1}$ with $\Delta_{\pm 1}=1/15$.  As previously said,
the case $N=2$ and $N=3$ are special because they can be identified with the
minimal model $M_3$ and $M_5$. The operators $\Phi^{\pm 1/2}$ and $\Phi^{\pm
  1}$ are identified respectively to the operators $\phi_{1,2}$ and
$\phi_{2,3}$ of the corresponding $M_3$ and $M_5$ Kac tables, see
Eq.(\ref{kac_minimal}).  As we will show more in detail below, the
representation module corresponding to $\Phi^{\pm 1}$ contains one descendant
state, $\varepsilon^{(N=3)}=A^{-1}_{-1/3}\Phi^{+1}=A^{1}_{-1/3}\Phi^{-1}$
singlet under $Z_3$ transformation (i.e. $q=0$). This singlet (energy)
operator has dimension $\Delta_{\varepsilon}=1/15+1/3=2/5$, it is a primary of
the Virasoro algebra and can be identified with the operator
$\varepsilon^{(N=3)}=\phi_{2,1}$ of the $M_5$ Kac table.

\subsection{Representation space: the disorder sector.}
\label{Rsector}
A key observation is that the theory we are considering is actually invariant
under the dihedral group $D_N$ which includes $Z_N$ as a subgroup.  This can
directly be seen from the symmetry of Eq.(\ref{opepara1})-(\ref{struct}) under
the conjugation of the $Z_N$ charge, $q\to N-q$. At the level of the lattice
model (see \ref{lattice model}), this comes from the invariance of the Hamiltonian under the
transformation $\sigma_k(x)\to \sigma_k^{+}(x)$.

The space of representation thus includes the $N-$plet of $Z_2$
operators which we denote as:
\begin{eqnarray}
 \{R_a(z,\bar{z}),\quad a=1,\ldots,N\}.
\end{eqnarray}
In the following, we refer to these operators as disorder operators.

The theory of disorder operators has been fully developed in \cite{Zamo2}.
A detailed discussion about the general properties (products, analytic
continuations) of these operators is given in \cite{Raoul1}.

The most important feature of the disorder operators is their non-abelian monodromy
with respect to the chiral fields $\{\Psi^{\pm k}\}$. This amounts to the
decomposition of the local products $\Psi^{k}(z) R_{a}(0)$ into half-integer
powers of $z$: 
\begin{equation}
 \Psi^{k}(z)R_{a}(0) =
\sum_{n}\frac{1}{(z)^{\Delta_{k}^{\Psi}+\frac{n}{2}}}\, A^{k}_{\frac{n}{2}}R_{a}(0),
\qquad k=1,2,\ldots,\lfloor N/2 \rfloor. \label{mode1R} 
\end{equation} 
The expansion of the
product $\Psi^{-k}(z)R_{a}(0)$ (with $k=1,2,\ldots,\lfloor N/2 \rfloor$) can be
obtained by an analytic continuation of $z$ around 0 on both sides of
Eq.~(\ref{mode1R}). The result is: 
\begin{equation}
 \Psi^{-k}(z)R_{a}(0) =
\sum_{n}\frac{(-1)^{n}}{(z)^{\Delta_{k}^{\Psi}+\frac{n}{2}}} A^{k}_{\frac{n}{2}} \,
R_{a-k}(0), \qquad k=1,2,\ldots,\lfloor N/2 \rfloor.  
\label{rexpansion}
\end{equation}

In accordance with these expansions, the mode operators
$A^{k}_{\frac{n}{2}}$ can be defined by the contour integrals
\begin{equation}
 A^{k}_{\frac{n}{2}}R_{a}(0)=\frac{1}{4\pi i}\oint_{C_{0}}
 {\rm d}z \, (z)^{\Delta_{k}^{\Psi}+\frac{n}{2}-1}\Psi^{k}(z)R_{a}(0),
\label{rmodes}
\end{equation}
where the integrations are defined by letting $z$ turn twice around the
operator $R_{a}(0)$ at the origin, exactly as described in \cite{Zamo2,Raoul1}.  The
commutation relations between the current modes in this sector can be computed
and the dimension of the primary disorder operators is (see Appendix
\ref{rsector_kac}):
\begin{equation}
\Delta^{R}_{s}=\frac{N-2+(N-2 s)^2}{16(N+2)} \quad s=0,1,\cdots, \lfloor N/2
\rfloor. 
\label{disorder_kac}
\end{equation}
Note that for $N=2$ one has the operators $R_{a}^{(0)}$ and $R_{a}^{(1)}$ with
dimension $\Delta^{R}_{0}=1/16$ and $\Delta^{R}_{1}=0$. This is of course expected as
the dihedral group $D_2=Z_2$, i.e. the cyclic $Z_N$ elements and the
reflection $Z_2$ elements of the group $D_N$ coincide for $N=2$. In particular
one has $R_{a}^{0}=\Phi^{\pm 1/2}$ while $R_{a}^{1}$ coincides with the
identity.  For $N=3$, the disorder sector contains the operator $R_{a}^{0}$
($\Delta^{R}_{0}=1/8$) and $R_{a}^{1}$ ($\Delta^{R}_{1}=1/40$). In terms of primaries
of the Virasoro algebra they coincide respectively with the $\phi_{1,2}$ and
$\phi_{2,2}$ operators of the $M_5$ Kac table, see again
Eq.(\ref{kac_minimal}).

\subsection{Stochastic motion in the internal space.}
\label{sle_zn}

We discuss now a possible SLE approach to describe
parafermionic theories.  In the previous section we have seen that the action
(\ref{modes}) and (\ref{rmodes}) of the chiral currents modes on the
representation fields amounts to a twist either of the $Z_N$ charge $q$ for
fields $\Phi^{q}(z)$ belonging to the $Z_N$ sector, or of the $Z_2$ index
$a$ for fields $R_{a}(z)$ in the disorder sector.  The b.c.c. 
operator is expected to transform as an operator in the $Z_N$ or in the $Z_2$
sector. By analogy to the case of the $SU(2)_N$ WZW \cite{Ludwig}, we
consider an independent stochastic motion in the additional degrees of freedom.
We provide a parafermionic field the following evolution:
\begin{equation}
X(z)\to X(z)+\frac{1}{2\pi i}\oint_{\mathcal{C}_z} d\, g_t(w) T(w)
X(z)+\frac{1}{2\pi i}\oint_{\mathcal{C}_z}d\, w (d \theta)^{k} (w)\Psi^{k}(w)
X(z),
\label{evolution}
\end{equation}
where $X(z)=\Phi^{q}$ or $X(z)=R_a(z)$ and the contour $\mathcal{C}_z$ is a
general closed contour around $z$ (remember that in the $R$ sector the
Riemann surface is two-fold). This evolution contains, in addition to the variation under an
infinitesimal conformal transformation $d g_t(w)$, a component with shifted
$Z_N$ charge $q\to q+k$ or $Z_2$ charge, $a \to a+k$. The (\ref{evolution})
can be seen as a motion in the parafermionic representation module.
In the sequel, we shall concentrate on considering only
the action of the most fundamental $\Psi^{\pm 1}$ fields.

The definition of the internal stochastic motion is given in such a way that
the corresponding Ito derivative of  $\psi(z)$ takes the
form (\ref{psi_dynamic}) or (\ref{wzw_psi_dinamic}). More precisely, we want
to derive a martingale condition as a linear relation between descendants at
the level two of some representation module. This allows for a direct
comparison with the CFT results.

In the case the boundary operator $\psi(z)$
coincides with a primary $\Phi^{q}$ in the $Z_N$ sector, we define:
\begin{equation}
(d \theta)_t^{k} (z)=\frac{(d \theta)^{k}_t}{(g_t(z)-\xi_t)^{-\Delta_{k}^{\Psi}+\delta^{q}_{k+q}+1}}, \quad \mbox{{\bf
    E}}[\theta^{k}_t\theta^{k'}_s]=\tau\,\delta^{k,-k'}\,\mbox{min}(s,t). 
\label{tau_sle_para_zn}
\end{equation}
The exponent in the denominator has been chosen according to
Eq.(\ref{eq3}) which encodes the OPE between the currents $\Psi^{k}$
and the fields $\Phi^{q}$.  Note that this evolution is very similar to
(\ref{tau_sle}) in the $su(2)$ Lie algebra. This is somehow expected since the
$Z(N)$ parafermionic theory corresponds to the coset $SU(2)_N/U(1)$ (see
appendix \ref{wzw_para}). 

In the case  when $\psi(z)$ transforms as a disorder $R_{a}$ field, we
define, according to Eq.(\ref{mode1R}):
\begin{equation}
(d \theta)_t^{k} (z)=\frac{(d \theta)^{k}_t}{(g_t(z)-\xi_t)^{-\Delta_{k}^{\Psi}+2}}, \quad \mbox{{\bf
    E}}[\theta^{k}_t\theta^{k'}_s]=\tau\,\delta^{k,-k'}\,\mbox{min}(s,t). 
\label{tau_sle_para_r}
\end{equation}

We point out that the definition of the stochastic motion given above is by
now purely algebraic and based on a formal analogy with the case $SU(2)$.
However, the CFT predictions and the identification of the possible interface
on the lattice model motivate the interest of such construction.

\subsection{Martingale conditions and parafermionic results.}
\label{mart_cond}

As we will discuss in some detail later, the Ising model ($N=2$ and $c=1/2$)
shows in its FK representation (see \ref{lattice model}) a domain wall described by the
SLE$_{16/3}$. The b.c.c operator creating such interface
is identified with the operator $\phi_{1,2}$ of the $M_3$ Kac table,
what is consistent with the Eq.(\ref{sle_cft_minimal}) and Eq.(\ref{sle_cft_bcc}). In
the parafermionic representations, we have seen that this operator corresponds
to the disorder $R_{a}^{(0)}$ operator or, equivantely, to the $\Phi^{\pm
  1/2}$ operator (remember that $N=2$ is special as $D_2$=$Z_2$). In the three
state Potts model ($N=3$), the operator $\phi_{1,2}$ of the $M_5$ Kac table is
again identified again with the disorder $R_{a}^{(0)}$ and it is expected to
create an SLE$_{24/5}$, in agreement again with Eq.(\ref{sle_cft_minimal}) and
Eq.(\ref{sle_cft_bcc}). The operator $\phi_{2,1}$, on the other
hand, can be associated to an interface described by the SLE$_{10/3}$ and it
corresponds to the
$\varepsilon^{(N=3)}=A^{1}_{-1/3}\Phi^{-1}=A^{-1}_{-1/3}\Phi^{1}$ operator
mentioned before.

On the basis of these correspondences, it is then natural to identify for
general $N$ the b.c.c.  operator $\psi (z)$ with the disorder
primary operator $R_{a}^{(0)}$ with dimension $\Delta_{0}=(N^2+N-2)/(16(N+2))$
and with the $Z_N$ singlet (i.e.  charge $q=0$) operator
$\varepsilon^{(N)}=A^{-1}_{-1/N}\Phi^{1}=A^{1}_{-1/N}\Phi^{-1}$ with dimension
$\Delta_{\varepsilon}=2/(N+2)$ . Note that this singlet operator is not a primary of the
parafermionic algebra but it appears as  descendant in the module of
$\Phi^{\pm 1}$ (see appendix \ref{nullvector}). 

Using Eq.(\ref{chidim}), Eq.(\ref{level_structure}) and Eq.(\ref{evolution}),
the stochastic motions (\ref{tau_sle_para_zn}) and (\ref{tau_sle_para_r})
amounts respectively to the following dynamics for $\psi(z)$ in the two cases
when $\psi(z) \equiv \psi_{\varepsilon}=\varepsilon(z)$ and $\psi(z) \equiv \psi_{R}=R_{a}^{(0)}$:
\begin{eqnarray}
d( \psi_{\varepsilon} (\xi_t))= d \xi_t L_{-1}\psi_{\varepsilon}(\xi_t)  + (d \theta)^{1}
A^{1}_{1/N-1} \psi_{\varepsilon}(\xi_t)+ (d \theta)^{-1}
A^{-1}_{1/N-1} \psi(\xi_t)_{\varepsilon}\nonumber \\
+ dt\left( \frac{\kappa}{2}
L_{-1}^2 +\frac{\tau}{2} \left[A^{1}_{-1/N-1}A^{-1}_{1/N-1}+A^{-1}_{-1/N-1}A^{1}_{1/N-1}\right]\right)\psi_{\varepsilon}(\xi_t).
\label{para_psi_dinamic_zn}
\end{eqnarray}
and $\psi(z) \equiv \psi_{R}=R_{a}^{(0)}$:
\begin{eqnarray}
  d( \psi_{R} (\xi_t))= d \xi_t L_{-1}\psi_{R}(\xi_t)  + (d \theta)^{1}
  A^{1}_{-1} \psi_{R}(\xi_t)+ (d \theta)^{-1}
  A^{-1}_{-1} \psi_{R}(\xi_t)\nonumber \\
  + dt\left( \frac{\kappa}{2}
    L_{-1}^2 +\frac{\tau}{2} \left[A^{1}_{-1}A^{-1}_{-1}+A^{-1}_{-1}A^{1}_{-1}\right]\right)\psi_{R}(\xi_t).
\label{para_psi_dinamic_R}
\end{eqnarray}
The martingale condition, to be compared with Eq.(\ref{sle_nul_vect}) and
Eq.(\ref{sle_nu_vect_wzw}), now reads:
\begin{eqnarray}
  \left(L_{-2}-\frac{\kappa}{4}
  L_{-1}^2 -\frac{\tau}{4}
  \left[A^{1}_{-1/N-1}A^{-1}_{1/N-1}+A^{-1}_{-1/N-1}A^{1}_{1/N-1}\right]\right)\psi_{\varepsilon}(\xi_t)&=&0 
\label{sle_nul_vect_para_1}
\\
\left(L_{-2}- \frac{\kappa}{4}
    L_{-1}^2 -\frac{\tau}{4} \left[A^{1}_{-1}A^{-1}_{-1}+A^{-1}_{-1}A^{1}_{-1}\right]\right)\psi_{R}(\xi_t)&=&0.
\label{sle_nul_vect_para_2}
\end{eqnarray}

We can now enounce the main results of this paper. We have studied the
representation modules of the $\varepsilon$ and $R_{a}^{(0)}$ operator. For  $N\geq 4$ one finds that there are at least two independent states at the second level of these operators.  We
have found the following relations at the second level of these modules:
\begin{eqnarray}
  \left(L_{-2}-\frac{N+1}{N+2}
  L_{-1}^2 -\frac{N^3}{(N+2)^3}
  \left[A^{1}_{-1/N-1}A^{-1}_{1/N-1}+A^{-1}_{-1/N-1}A^{1}_{1/N-1}\right]\right)\psi_{\varepsilon}(\xi_t)&=&0 \label{para_nul_vect_1}\\
\left(L_{-2}- \frac{N+2}{N+1}
    L_{-1}^2 -\frac{ 2^{-4/N} N^3}{4(N+2)(N+1)}\left[A^{1}_{-1}A^{-1}_{-1}+A^{-1}_{-1}A^{1}_{-1}\right]\right)\psi_{R}(\xi_t)&=&0,
\label{para_nul_vect_2}
\end{eqnarray}
The derivation of these relations is shown in
appendix \ref{nullvector}. We have verified that, for $N=2$, the relation
(\ref{para_nul_vect_2}) reduces to (\ref{vir_nul_vect_1}) and for $N=3$, the
Eq.(\ref{para_nul_vect_1}) and Eq.(\ref{para_nul_vect_2}) are respectively
equivalent to Eq.(\ref{vir_nul_vect_2}) and Eq.(\ref{vir_nul_vect_1}). In
these cases, which as already said coincide with minimal models $M_3$ and
$M_5$, all the descendants of the theory can be obtained by applying negative
Virasoro modes. In our case, this amounts to the fact that the states at the
second level constructed from the application of parafermionic modes can be
written as a linear combination of $L_{-1}^2$ and $L_{-2}$.

Comparing Eq.(\ref{sle_nul_vect_para_1}) and Eq.(\ref{sle_nul_vect_para_2}) to
Eq.(\ref{para_nul_vect_1}) and Eq.(\ref{para_nul_vect_1}), we obtain, for
$N\geq 4$ and  for the
two SLEs evolutions:
\begin{eqnarray}
\kappa_1&=&4 \frac{N+1}{N+2}\quad \tau_1= 4\frac{N^3}{(N+2)^3} \label{Sle1} \\
\kappa_2&=&4 \frac{N+2}{N+1}\quad \tau_2= 2^{-4/N}\frac{N^3}{(N+2)(N+1)}\label{Sle2}
\end{eqnarray}

\section{$Z(N)$  self-dual critical  spin models.} 
\label{lattice model}

\begin{figure}
\begin{center}
 \leavevmode
\includegraphics[scale=0.5]{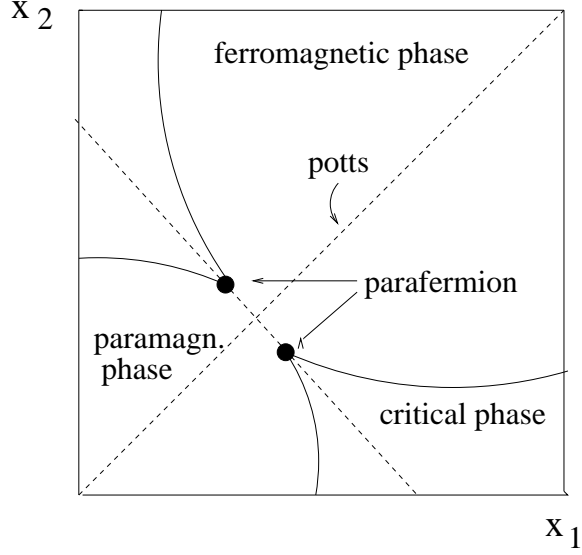}
 \end{center}
 \protect\caption[3]{\label{diaZ5}Phase diagram of th $Z(5)$ spin model}
\end{figure}

Consider a square lattice with the spin variables $\sigma_{j}$ at each
sites $j$ taking $N$ possible values:
\begin{equation}
\sigma_{j}=\exp\left[\frac{i2\pi}{N}n(j)\right] \quad n(j)=0,1,\cdots,N-1.
\label{spin_variables}
\end{equation}
The most general $Z_N$-invariant spin model  with nearest-neighbor
interactions is defined by the reduced Hamiltonian \cite{Zama_lat,Dotsi_lat,Fradkin_lat}:
\begin{equation}
H[n]=-\sum_{m=1}^{\lfloor N/2 \rfloor} J_{m}\left[\cos \left(\frac{2\pi m n}{N}\right)-1\right].
\label{reducedH}
\end{equation}
and the associated partition function reads:
\begin{equation}
Z=\sum_{\{\sigma\}}\exp\left[-\beta \sum_{<ij>} H[n(i)-n(j)]\right]
\label{partition1}
\end{equation}
In fact, the Potts model is recovered in the case 
$J_m=J$ for all $m$. The model presents, in this case, a permutational $S_N$
symmetry. Another meaningful model is the clock model which is defined by the
partition function 
(\ref{partition1}) with $J_m=J\delta_{m,1}$. 
Defining the Boltzmann weights:
\begin{equation}
x_n=\exp\left[-\beta H(n)\right], \quad n=0,1,\cdots,N-1,
\end{equation}
the general $Z(N)$ spin system is described by $\lfloor N/2\rfloor$ independent
parameters $x$. The Kramers and Wannier (order-disorder) duality has proven to
be a powerful tool for examining the behavior of these models. The phase
diagram of the model (\ref{partition1}) for $N=5$ states of spin is shown in
Fig.(\ref{diaZ5}). It has been show that the $Z(5)$ model is self-dual on the
line $x_1+x_2=1/2(\sqrt{5}-1)$. Here we are interested in the
existence of two critical points $(x_1^{*}, x_2^{*})$ and $(x_2^{*},
x_1^{*})$, where $x_1^{*}=\sin(\pi/20)/\sin(3\pi/20)\approx 0.34$ and
$x_2^{*}=\sin(\pi/4)/\sin(7\pi/20)x_1^{*}\approx 0.27$, where the model is
completely integrable.  The $Z(5)$ parafermionic theory defined in the previous
sections describes the continuum limit of the lattice model at these points.
In general, in the self-dual subspace of (\ref{reducedH})-(\ref{partition1}),
which contains also the Potts model the clock model, the $Z(N)$ spin model is
completely integrable at the points \cite{Zama_lat2} :
\begin{eqnarray}
x^{*}_0&=& 1 \nonumber \\
x^{*}_n&=& \prod_{k=0}^{n-1} \sin \left(\frac{\pi k}{N}+\frac{\pi}{4
    N}\right)\left[\sin \left(\frac{\pi (k+1)}{N}-\frac{\pi}{4 N}\right)\right]^{-1}.
\label{integrablecond}
\end{eqnarray}
These critical points are described by the $Z(N)$ parafermionic field theories.

\subsection{ Boundary conditions and SLE interfaces.} 
\begin{figure}
\begin{center}   
 \includegraphics[scale=0.5]{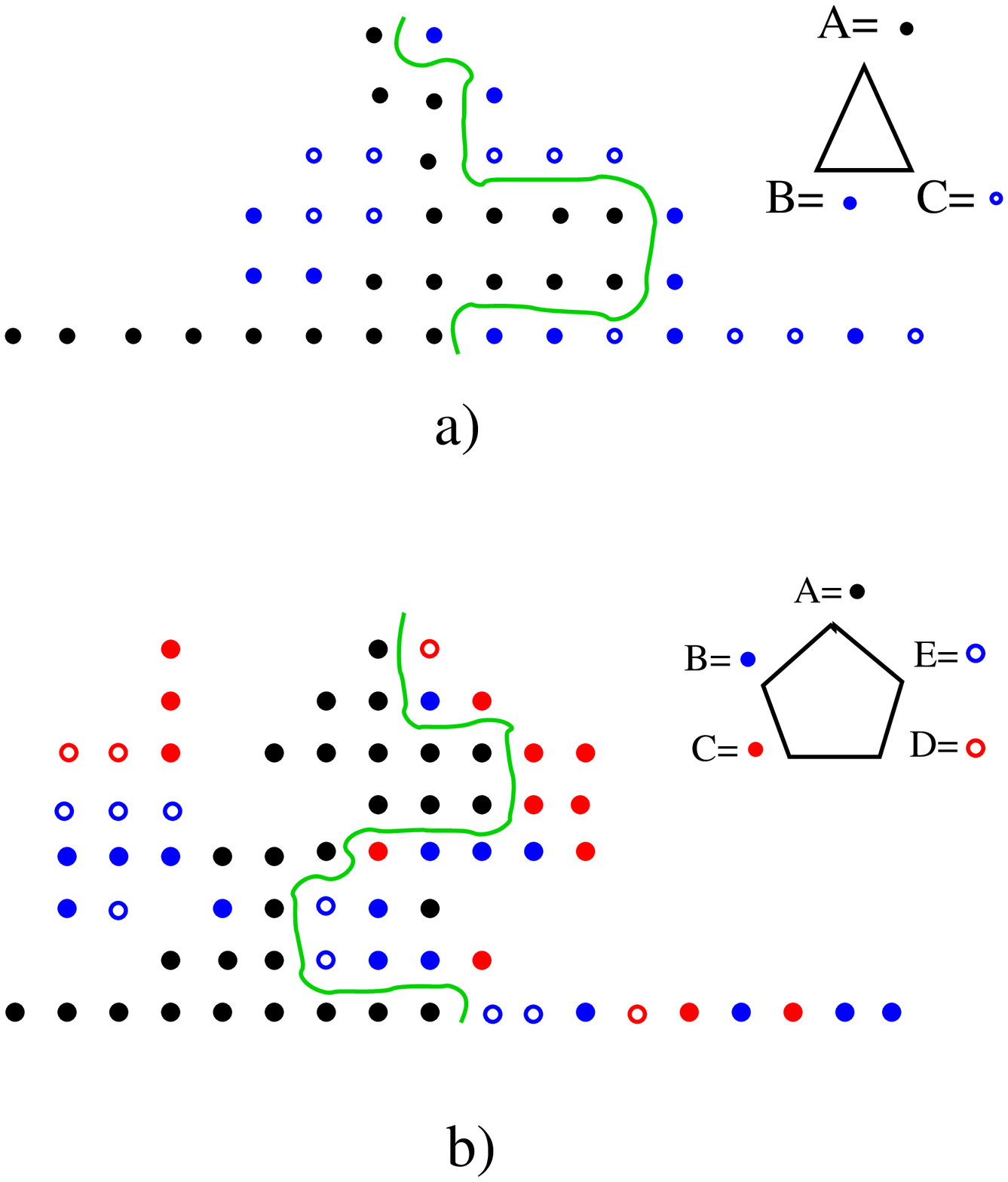}
 \end{center} 
 \protect\caption[3]{\label{slekl4}a)Three state Potts model: interface (green
   curve) between the $A$ spins (black circles) and the $B$ (full blue circles) and
   $C$ (empty blue circles). The boundary conditions $A|B+C$ as in the
   text. b) The interface defined in the $Z(5)$ spin model with boundary conditions
   $A|B+C+D+E$.}
\end{figure}

Consider the three-states Potts model.  From the Eq.(\ref{sle_cft_minimal})
and the Eq.(\ref{sle_cft_bcc}), as previously mentioned, the b.c.c. 
operator $\psi_{\varepsilon}^{(N=3)}$ is expected to create an 
SLE$_{10/3}$ trace. In this paragraph we suggest a definition of the
interfaces within the lattice models whose scaling limit is described in these
SLE processes.  In the following we
indicate the possible values of the spins with the letters $A, B\cdots$. The
b.c.c.  operator $\psi_{\varepsilon}^{(N=3)}$ has been shown to generate the
boundary condition where the spins are fixed to (say) the value $A$ on the
left side of the origin while can take the value $B$ or $C$ with equal
probability on the right side \cite{Cardy_bcft2}. With this boundary condition, to which we refer
with the short-hand notation $A|B+C$, there exists a single domain wall
between the spin $A$ and the spins $B$ and $C$, as shown in
Fig.\ref{slekl4}. The statistical properties of this interface has recently
been analyzed by numerical means in \cite{Cardy_potts} and are shown to be well
described in the continuum limit by an SLE$_{10/3}$.

The generalization to the $N$-states spin model is straightforward. Indeed,
one can show (see appendix \ref{boundarystates}) that the b.c.c. operator $\psi_{\varepsilon}^{(N)}$
produces the boundary conditions $A|B+C+D+\cdots$, where the spins on the left
of the origin take the value $A$ while the ones on the right take with equal
probability the $N-1$ values of spin different from $A$.  As in the case of
$N=3$, there exists a single boundary domain between the $A$ spin and the
$B,C,\cdots..$. Note that the boundary conditions do not satisfy the
reflection symmetry and the measure on the curve should reflect this
asymmetry.  However, in the continuum limit the reflection symmetry is
expected to be restored as explicitely shown in \cite{Cardy_potts}. This can
be argued from the fact that, in this limit, the statistical expectation values
can be expressed in terms of CFT correlation functions which contain the
b.c.c.  operators, see Eq.(\ref{martingale_cft}). These correlation functions
satisfy indeed the reflection symmetry.

It is interesting to note that one can assign a ``color'' to the interface
depending on the values of the spins ($B,C,\cdots$) at the right of the domain wall.  The
point is that, for $N<4$, the Boltzmann weights $x_n^{*}$,
$n=1,2,\cdots,N-1$, are equal: the energy of one bond does not depend on the
difference between the values of the nearest-neighbor spins. In this sense,
the $Z_N$ symmetry is trivially realized and this is essentially the reason
why, for $N=2,3$, the correspondent CFT can be constructed from the conformal
invariance alone. However this is not true for $N\geq 4$ where the Boltzmann weights
$x_n^{*}$, $n=1,2,\cdots \lfloor N/2 \rfloor$ are different.  In the continuum
limit, this amounts to a more structured chiral algebra with a central charge
$c\geq1$.  From the point of view of the lattice model, one can observe that for $N=2,3$ the
energy cost to have a spin $B$ or $C$ at one side of the interface is the
same. Thus, as $N\geq 4$, one expects the internal spin degree of freedom
to play an important role in the spatial evolution of the interface.  This is
somehow in agreement with the need to introduce an additional stochastic
motion in the $Z_N$ group. However a clear the geometrical interpretation of the
stochastic evolution in the representation modules (see section \ref{sle_zn})
is missing. 

From the results (\ref{para_nul_vect_1}) and (\ref{Sle1}), we hypothesize that
the geometric properties of this interface for $N\geq 4$ are described by a
$SLE$ with $\kappa= 4(N+1)/(N+2)$. In particular, since $\kappa\leq 4$,
we expect the domain under consideration to be a simple curve for each $N$.

We discuss now the identification of the interface on the lattice associated
to the b.c.c operator $\psi_{R}$, see Eq.(\ref{para_nul_vect_2}).  In the
Ising and three states Potts model, this operator is associated to an
SLE$_{16/3}$ and SLE$_{24/5}$ respectively. The correspondent interfaces on
the lattice can be identified with the help of the Fortuin-Kasteleyn (FK)
representations.

The partition function (\ref{partition1}) for the $N=2$ and $N=3$ critical
models take the form:
\begin{equation}
Z=\sum_{\{ \sigma\}} \exp \left[ -\beta \sum_{<ij>} H(n(i)-n(j))\right]=\sum_{\{
  \sigma\}} \prod_{<ij>}\left[x_1^{*}+(1-x_1^{*})\delta_{n(i),n(j)}\right],
\label{fk1}
\end{equation}
where $x_1^{*}=\sqrt{2} -1$ or $x_1^{*}=x_{-1}^{*}=(\sqrt{3} -1)/2$ is the
critical point of the Ising and of the three states Potts model respectively.
By expanding the product (\ref{fk1}) and summing over all the spin
configurations, one obtains the FK random cluster representation:
\begin{equation}
Z=\sum_{\mathcal{C}} (x_1^{*})^{M-b}(1-x_1^{*})^{b} N^{c}
\label{fk2}
\end{equation}
where the sum is over the subgraphs $\mathcal{C}$ of $\mathcal{G}$,
$\mathcal{C}\subseteq \mathcal{G}$, the graph $\mathcal{G}=(V,E)$ being
composed by the $V$ sites and the $E$ edges of the square lattice. Each graph
$\mathcal{C}$ is specified by the $b$ bonds on the edges and the $c$ clusters
(=connected component which include the single sites) in a given bond
configuration.

Consider the following boundary conditions: all the edges on the negative axis
carry bonds while all the edges on the positive axis carry no bonds.  For the
spin variables this means that  all the spins located at
the left of the origin take the same value while the spins on right side are
unconstrained, i.e. they are free (F) to take all the $N$ possible values.
 
In such an arrangement, each graph $\mathcal{C}$ presents a cluster growing from the
negative axis into the upper half-plane. The
boundary of this cluster, starting from the origin, is conjectured (see for
istance \cite{Walter}, chapter 5) to be
statistically equivalent in the continuum limit to the SLE$_{16/3}$ for $N=2$ 
(Ising model) and to the SLE$_{24/5}$ for $N=3$ (the three states Potts
model).  This is in
agreement with the properties of the b.c.c. $\psi_{R}^{(N)}$ operators,
$N=2,3,4$, which have been shown \cite{Bauer_R} to be associated to the $A|F$ boundary
conditions.

The generalization to the case $N\geq 4$ is more cumbersome. To our knowledge,
the modular properties of the $R-$sector characters, and thus the boundary
conditions associated to these operators, have not been yet studied for these
cases. This motivates a more detailed analysis of the boundary conformal
properties of this sector. However, we point out that at the level of the
parafermionic algebra representations the main features of the $\psi_{R}$
fields, such as the structure and the degeneracies of the representation
module or the current zero modes eigenvalues (see Appendix \ref{rsector_kac}), directly
generalize to each $N\geq 2$.  The formula (\ref{disorder_kac}) is an example.
It is thus quite natural to suppose that the operators $\psi_{R}^{(N)}$
produce the $A|F$ boundary conditions for each $N$. Another observation is
that it is quite direct to find an alternative formulation of
(\ref{partition1}) in terms of colored clusters on a graph. For sake of
clarity, we focus our attention on the case $N=5$.

The partition function (\ref{partition1}) for $N=5$ at the critical
point $x_n^{*}$, see Eq.(\ref{integrablecond}), can be rewritten as:
\begin{equation}
Z^{(N=5)}=\sum_{\{\sigma\}}\prod_{<ij>} \left[x_1^{*}+(1-x_1^{*})\delta_{n(i),n(j)}+(x_2^{*}-x_1^{*})\delta_{n(i),n(j) \pm 2}\right].
\label{partition2}
\end{equation}
If $x_1^{*}=x_2^{*}$ one recovers the five states Potts model
in the usual FK representation.  

\begin{figure} 
\begin{center} 
 \leavevmode  
 \includegraphics[scale=0.5]{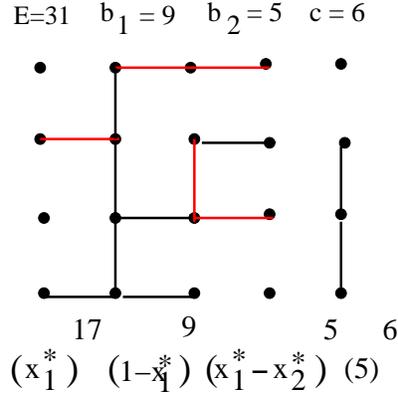}
 \end{center} 
 \protect\caption[3]{\label{fk_colorato}Cluster expansion of the $Z(5)$ spin  
   model. The spins at the vertices of a black bond have the same value while
   the ones at the vertices of a red bond  differ by an angle $4\pi/5$}
\end{figure}

Expanding the factors in (\ref{partition2}), one has to take into account two
type of bonds: the bonds (type 1) connecting spins which take equal values
and the bonds (type 2) connecting spins which differ by an angle $4\pi/5$.
Then, summing over the spin configurations, the partition function can be cast
in terms of a sum over ``colored'' graphs $\mathcal{C'}$:
\begin{equation}
  Z^{(N=5)}=\sum_{\mathcal{G'}} (x_1^{*})^{E-b_1-b_2}(1-x_1^{*})^{b_1}(x_2^{*}-x_1^{*})^{b_2}5^{c}.
\label{z5_fk}
\end{equation}
Each graph $\mathcal{C'}$ is specified by the number $c$ of connected
component and by the numbers $b_1$ and $b_2$ of the bonds of type 1 and type 2.
Naturally, the sum over the differences of
the spins along a loop must be zero (modulo $2\pi$).  For instance, a loop
formed by three bonds of type 2 and one bond of type 1 is not permitted. The
prime in $\mathcal{C'}$ indicates that the sum in (\ref{z5_fk}) is taken over
the allowed graphs. A typical configuration is shown in
Fig.(\ref{fk_colorato}).

\begin{figure} 
\begin{center} 
 \leavevmode  
 \includegraphics[scale=0.5]{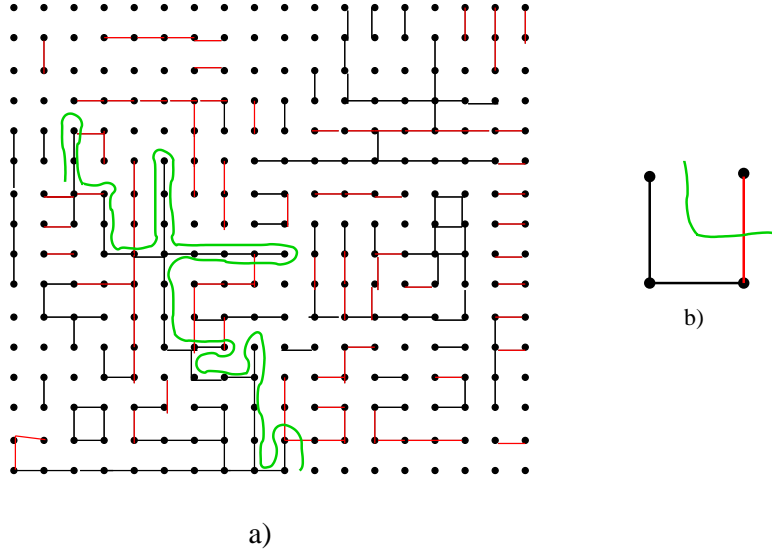}
 \end{center} 
 \protect\caption[3]{\label{slekp4} a) A typical cluster configuration of
   the $Z(5)$ spin model with type 1 (black) and type 2 (red) bonds. The
   boundary conditions on the real axes are as in the text. The green curve is the boundary of
   the cluster growing from the negative axes. b) A piece of the interface
   which crosses red or empty bond.}
\end{figure}

By analogy to the Ising and three states Potts model, we consider the boundary
conditions in which the spins have a fixed value on the negative axis while
they are free on the positive axis. With such boundary conditions,
there is a cluster of bonds of type $1$ growing from the negative axis, see
Fig.(\ref{slekp4}) . 

From (\ref{Sle2}), we conjecture the geometric properties of the boundary of this cluster are
described in the continuum limit by the SLE$_{14/3}$. 

The above discussion can be directly generalized to each value of $N$. In
particular, the $Z(N)$ spin model admits a random cluster representations
with $\lfloor (N-1)/2 \rfloor$ types of bonds. Imposing the conditions on the
real axis discussed above, we conjecture that the measure of the boundary of the
cluster growing from the negative real axis is described by an SLE with $\kappa=4(N+2)/(N+1)$.

Finally we remark that, during its evolution, the interface crosses edges
without  bonds or with bonds of the type $2,3,\cdots \lfloor (N-1)/2 \rfloor$ which
have different energetic costs.  Analogously to the case of the interface
generated by the $A|B+C+\cdots$ boundary conditions, one can color the interface
depending on the type of bond it crosses.  Again, the interface is naturally
provided of an $Z_N$ additional internal degree of freedom, in agreement with
the approach proposed in Section (\ref{sle_zn}).

\section{Conclusions.}
\label{conclusion}
In this paper we have considered the possibility to extend an SLE approach to
CFTs with additional $Z_N$ discrete symmetries.  These theories describe the
continuum limit of the self-dual critical $Z(N)$ spin models. The case $N=2$
and $N=3$ correspond to the Ising and three states Potts model where the
identification of the interface on the lattice and of the correspondent b.c.c.
operators are known. Using these results, we have identified the possible
b.c.c. operators for general $N$. We show that these operators satisfy a two
level null vector condition. For $N\geq 4$, an additional term enter these
relations. This term is obtained from the action of the parafermionic currents
and, for $N\geq 4$ ($c\geq 1$), it cannot be derived by the Virasoro modes
action. These results led us to propose an additional stochastic motion in the
different $Z_N$ charge sectors and in the $Z_2$ sectors which describe the
representation modules of the parafermionic algebra. We have finally discussed
the possible interfaces in the lattice to which the CFT/SLE results should
apply. The fact that SLE candidates have been identified in lattice spin
models described by non-conformal minimal theories opens the possibility to
study the SLE/non-minimal CFT connection by a numerical approach. 

It will be interesting to verify our predictions by numerical studies. Further
we believe these studies can also better clarify the geometrical
interpretation of the additional stochastic motion in the internal space.

{\bf Acknowledgments:} We would like to thank M.~ Bauer, J.~Cardy, I.~Gruzberg, C.~Hagendorf,
A.~Ludwig, V.~Riva for very helpful discussions.  We are particularly grateful
to D.~Bernard and P.~Le Doussal for very useful advises and support all along
this work.
\appendix
\section{Relation between the  $SU(2)_N$ WZW model and the $Z(N)$ parafermionic theory.}
\label{wzw_para}
In this appendix we briefly show how the stochastic motion (\ref{tau_sle})
for the $SU(2)_N$ WZW theories can be somehow compared to the one
(\ref{tau_sle_para_zn}) by taking into account that the $Z(N)$ parafermionic theory corresponds to the coset $SU(2)_N/U(1)$.
The $SU(2)_N$ WZW theory can then be seen as a combined system of the $Z_N$
parafermionic theory and one free massless boson field $\phi(z)$. The chiral
currents generating the Kac-Moody algebra (\ref{kac_moody}) can be written in
terms of the parafermionic currents and the free field as: 
\begin{eqnarray}
J^{\pm}&=& \sqrt{N}\Psi^{\pm 1}(z)\exp\left[\pm i \frac{1}{\sqrt{N}}\phi(z)\right] \nonumber
\\
J^{0}&=&i \frac{\sqrt{N}}{2} \partial \phi(z). 
\label{curr_wzw_para}
\end{eqnarray}
The stress-energy tensor $T^{SU(2)_N}$ and $T^{Z(N)}$ of the two theories are
symply related by $T^{SU(2)_N}=T^{Z(N)}+T^{U(1)}$, where $T^{U(1)}=-1/4
(\partial \phi)^2$ is the stress-energy tensor of the free bosonic theory.

All the primary fields $\phi_{j}$ of the $WZW$ theory can be obtained by a
product of parafermionic fields in the $Z_N$ sector and of vertex operators of
the bosonic theory:
\begin{equation}
\phi_{j}(z)=\Phi^{j}\exp\left[i \frac{j}{\sqrt{N}} \phi(z)\right].
\label{su2para1}
\end{equation} 
Using the fact that the conformal dimension $\Delta^{vert.}_{\alpha}$ of the
vertex operator $\exp[\pm i \alpha \phi(z)/\sqrt{N}]$ is
$\Delta^{vert.}_{\alpha}=\alpha^2/N$, one can easily verify that
$\Delta_{j}=\Delta^{\Phi}_{j}+ \Delta^{vert.}_{j}$, see Eq. (\ref{dim_wzw})
and Eq (\ref{kac_zn}).  Note that the field $\phi_{j}(z)$ in
Eq.(\ref{su2para1}) satisfies $J^{0}_{0}\phi_{j}=j \phi^{m}_{j}$. In the
following we denote $\phi^{m}_{j}(z)$, with $m=-j, -j+1,\cdots,j$, the field
in the $SU(2)$ representation of spin $j$ such that $J^{0}_{0}\phi^{m}_{j}=m
\phi^{m}_{j}$. The field $\phi^{j-k}_{j}$ can be expressed in
terms of the parafermionic primaries $\Phi^{q}$ and of vertex operators as:
\begin{equation}
\phi^{j-k}_{j}(z)=(A^{-k}_{-\delta^{j}_{j-k}}\Phi^{j})\exp\left[i \frac{j-k}{\sqrt{N}} \phi(z)\right].
\label{su2para2}
\end{equation}
where $\delta^{q}_{k}$ has been defined in Eq.(\ref{level_structure}). For
instance, the $SU(2)$ operator $\phi_{1}^{0}$  corresponds to
$\varepsilon^{(N)}$, see section \ref{mart_cond}, while the operator $\phi_{\pm
  1}^{1}$ to $\Phi^{\pm 1}\exp[\pm i \phi(z)]$

Using the above relations, one can compare the definitions 
(\ref{tau_sle}) and (\ref{tau_sle_para_zn}), and in particular the exponent in
the denominator of the two formulas. This exponent actually determines
the current mode acting on the b.c.c. field under his evolution. In the
$SU(2)_N$ theory this is the $J^{a}_{-1}$ mode, as it can be seen in Eq.(\ref{wzw_psi_dinamic}).
Let us consider for istance the state $J^{-}_{-1}\phi^{1}_{1}$:
\begin{eqnarray}
J^{-}_{-1}\phi^{1}_{1}&=&1/(2\pi i)\oint_{w} d z
\,(z-w)^{-1} \,J^{-}(z)\phi^{1}_{1}(w) \nonumber \\ 
&&\propto 1/(2\pi i)\oint_{w} d z (z-w)^{-1}
\left(\Psi^{-1}(z) \Phi^{1}(w)\right)\left(\exp\left[\frac{-i}{\sqrt{N}} \phi(z)\right]\exp\left[\frac{i}{\sqrt{N}} \phi(w)\right]\right) \nonumber \\
&\propto& 1/(2\pi i)\oint_{w} d z  (z-w)^{-1-2/N}\left(\Psi^{-1}(z)
\Phi^{1}(w)\right) \left( 1+\frac{i}{\sqrt{N}}(z-w) \partial \phi(w) +\cdots \right),
\end{eqnarray}
where we have used the relations (\ref{su2para1}) and the divergences coming from the OPE in the
bosonic sector:
\begin{equation}
\exp\left[i \frac{q}{\sqrt{N}} \phi(z)\right]\exp\left[i
  \frac{q'}{\sqrt{N}} \phi(z')\right]=(z-z')^{2 \frac{q \, q'}{N}}\exp\left[ i \frac{q+q'}{\sqrt{N}} \phi(z')\right]+\cdots
\label{ope_bosonic}
\end{equation}
The above expression gives precisely the state $A^{-1}_{-1/N-1}\Phi^{1}$
multiplied by a bosonic piece. In the same way one can show that the term
$J^{-}_{-1}\phi^{0}_{1}$ and the term 
$(J^{+}_{-1}J^{-}_{-1}+J^{-}_{-1}J^{+}_{-1})\phi^{0}_{1}$ correspond
respectively to $A^{-1}_{1/N-1}\varepsilon^{(N)}$ and to 
$(A^{1}_{-1/N-1}A^{-1}_{1/N-1}+A^{-1}_{-1/N-1}A^{1}_{1/N-1})\varepsilon^{(N)}$
which also appear in the Ito derivative of the $Z(N)$ b.c.c. parafermionic
operator, see Eq.(\ref{para_psi_dinamic_zn}).

\section{Null vectors in the parafermionic modules.}
\label{nullvector}

\subsection{Commutation relations of parafermionic current modes:$Z_N$ sector}
\label{commutation_relation_zn}

The commutation relations of the mode operators can be deduced from
Eq.(\ref{modes}) by using standard techniques in the complex plane. Consider
for instance the commutation relation of the modes of $\Psi^{1}(z)$ in the
$Z_N$ singlet sector $\Phi^{0}$. We use the short-hand notation
$\{\Psi^{1},\Psi^{1}\}\Phi^{0}$ to indicate such relations. One consider the
double integral: 
\begin{equation}
\frac{1}{2\pi i}\oint_{C_{0}}{\rm d}z_1 \oint_{C_{0}}{\rm d}\,z_2 \,z_1^{n}z_2^{m}
(z_1-z_2)^{2/N-1}\Psi^{1}(z_1) \Psi^{1}(z_2) \Phi^{0}(0),
\end{equation}
where one adds the term $(z_1-z_2)^{2/N-1}$ to make the integrand
single-valued. By using the Eqs.(\ref{opepara1})-(\ref{eq3}) and expanding in series the
term $(z_1-z_2)^{2/N-1}$, the Cauchy theorem gives the following relations for
$\{\Psi^{1},\Psi^{1}\}\Phi^{0}$:
\begin{equation}
\sum_{l} D^{l}_{(2-N)/N}\left[ A^{1}_{(3-N)/N+n}A^{1}_{1/N+m}+
  A^{1}_{(3-N)/N+m}A^{1}_{1/N+n}\right]\Phi^{0}=\lambda^{1,1}_{2} A^{2}_{(4-N)/N+n+m}\Phi^{0}.
\label{psipsi}
\end{equation}  
The coefficients $D^{l}_{\alpha}$ are defined from the development 
\begin{equation}
(1-x)^{\alpha}=\sum^{\infty}_{l=0}D^{l}_{\alpha}x^{l}.  
\end{equation} 
The relations $\{\Psi^{1}\Psi^{-1}\}\Phi^{1}$, for instance, are derived from
the double integral:
\begin{equation}
\frac{1}{2\pi i}\oint_{C_{0}}{\rm d}z_1 \oint_{C_{0}}{\rm d}\,z_2
\,z_1^{2/N+n}z_2^{(N-2)/N +m}
(z_1-z_2)^{-(N+2)/N}\Psi^{1}(z_1) \Psi^{-1}(z_2) \Phi^{1}(0).
\end{equation}
The exponent of the $(z_1-z_2)$ term has been chosen on the basis of the OPE
(\ref{opepara2}). In particular it allows the modes of the stress energy,
appearing at the second order in (\ref{opepara2}), to enter in the relation 
$\{\Psi^{1}\Psi^{-1}\}\Phi^{\pm 1}$. One obtains:
\begin{equation}
\sum_{l} D^{l}_{-(N-2)/N}\left[ A^{1}_{1/N+n-1}A^{-1}_{-1/N+m+1}+
  A^{-1}_{-3/N+m}A^{1}_{3/N+n}\right]\Phi^{\pm 1}=\left[\frac{1}{2}\left(\frac{2}{N}+n\right)\left(\frac{2}{N}+n-1\right)\delta_{n+m,0}+\frac{N+2}{N}L_{n+m}\right]\Phi^{\pm 1}.
\label{psivir}
\end{equation} 
From the above relations, and using the primary condition (\ref{eq3}), one can
readily obtain the conformal dimension $\Delta_{1}$ of a primary field
$\Phi^{\pm 1}$. Setting $n=m=0$ in Eq.(\ref{psivir}), one obtains
$\Delta^{\Phi}_{1}=(N-2)/(N(N+2))$.

For completeness, we give below in a compact form the 
the commutation relations $\{\Psi^{1},\Psi^{1}\}\Phi^{q}$ and
$\{\Psi^{1},\Psi^{-1}\}\Phi^{q}$ which determine the space of the representations :
\begin{eqnarray}
 &&\sum^{\infty}_{l=0}D^{l}_{(2-N)/N} \left(
 A^{1}_{-\delta^{q+1}_{q+2}+s(q)+n-l}
 A^{1}_{-\delta^{q}_{q+1}+m+l}+A^{1}_{-\delta^{q+1}_{q+1}+s(q)+m-l}
 A^{1}_{-\delta^{q}_{q+1}+n+l} \right) \Phi^{q} 
 =\lambda^{1,1}_{2}
 A^{2}_{-\delta^{q}_{q+2}+t(q)+n+m}\Phi^{q}
 \label{psi_2}\\
&&\sum^{\infty}_{l=0}D^{l}_{-(2+N)/N} \left(
 A^{1}_{-\delta^{q-1}_{q}+u(1,q)+n-l}
 A^{-1}_{-\delta^{q}_{q-1}+m+l}+A^{-1}_{-\delta^{q+1}_{q}+u(-1,q)+m-l}
 A^{1}_{-\delta^{q}_{q+1}+n+l} \right) \Phi^{q}=\nonumber\\
&&= \left[
\frac{1}{2}(\frac{2|q|}{N}+n)(\frac{2|q|}{N}+n-1)+\frac{N+2}{N}L_{n+m-1+u(1,q)}\right]\Phi^{q} \label{psi_vir}.
\end{eqnarray}
The integers $s(q)$, $t(q)$ and $u(\pm 1,q)$ shift the indeces of the
parafermionic modes:
\begin{eqnarray}
s(q)&=& \delta^{q+1}_{q+2}-\delta^{q}_{q+1}+\frac{2}{N}-1 \nonumber \\
t(q)&=&\delta^{q}_{q+2}-2\delta^{q}_{q+1}+\frac{2}{N} -1\nonumber \\
u(k,q)&=&  \delta^{q-k}_{q}-\delta^{q}_{q-k}-\frac{N+2}{N},
\end{eqnarray}
where $\delta^{q}_{q+2}$ is defined in Eq.(\ref{level_structure}).
Finally the
algebra is then completed by the commutators between the parafermions modes
$A^{\pm 1}$ and $L_{n}$:
\begin{equation}
 \left(A^{\pm 1}_{-\delta^{q}_{q\pm 1}+m}L_{n}-L_{n}A^{\pm
     1}_{-\delta^{q}_{q\pm 1}+m}\right)
 \Phi^{q}= \left[ (1-\Delta^{\Psi}_{1})n+m-\delta^{q}_{q \pm 1} \right]
 A^{\pm 1}_{-\delta^{q}_{q \pm 1}+m+n}\Phi^{q}. \label{parvira}.
\end{equation}
Analogously to what we have seen above for the special case of the $\Phi^{\pm
  1}$ field, the formula (\ref{kac_zn}) giving the dimension of the primary
fields is obtained by taking $n=m=0$ ($n=1,m=0$) in the Eq.(\ref{psi_vir}) for
$q\neq 0$ ($q=0$). 
\subsection{$\Phi^{\pm 1}$ module.}
\begin{figure}
\begin{center}
 \leavevmode
 \includegraphics[scale=0.5]{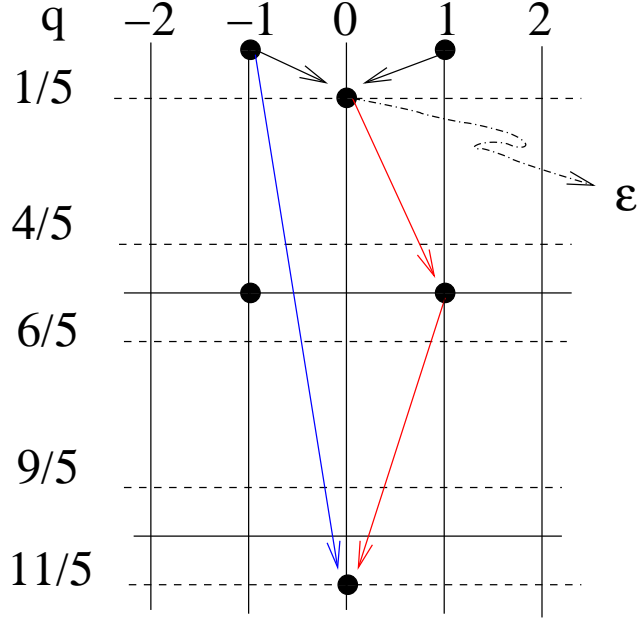}
 \end{center}
 \protect\caption[3]{\label{phi1modulo} Representation module of $\Phi^{\pm
     1}$ and the action of the parafermionic modes giving the operator
   $\varepsilon$ and its second level descendants}
\end{figure}

In the following we consider in more detail the module of the representation
$\Phi^{\pm1}$.
The structure of the
corresponding module is shown in Fig.(\ref{phi1modulo}) for the case $N=5$. 

 We have seen
above that the dimension of the operator are given by the commutation
relations Eq.(\ref{psi_vir}) without imposing any degeneracy in the module.
This in general is not the case, as it can be seen in the study of the
representation of the minimal models or of other parafermionic algebras.
However, in order to characterize completely the representation one needs to
fix the eigenvalue $h_{\pm 1}$ of the zero mode defined in
Eq.(\ref{zeroeigen}). We show this is easily obtained by imposing a degeneracy
at level $1/N$ where the first descendent states in this doublet are the two
singlet states $A^{-1}_{-1/N}\Phi^{+1}$ and $A^{1}_{-1/N}\Phi^{-1}$.
Forming the linear combination
\begin{equation}
 \chi^{0}_{-\frac{1}{N}}= a A^{1}_{-\frac{1}{N}}\Phi^{-1}+b A^{-1}_{-\frac
 {1}{N}}\Phi^{1},
 \label{singD1a}
\end{equation}
we wish to make it into a primary operator, i.e., to ensure that it is
annihilated upon action by positive index mode operators.
In this case it will be sufficient to verify that
\begin{equation}
 A^{1}_{+\frac{1}{N}}\chi^{0}_{-\frac{1}{N}}=0\quad\mbox{and}\quad
 A^{-1}_{+\frac{1}{N}}\chi^{0}_{-\frac{1}{N}}=0.
 \label{condD1a}
\end{equation} 
The degeneracy conditions $A^{\pm 1}_{1/N}\chi^0_{-1/N}=0$ are satisfied if 
\begin{equation}
 \left( \mu_{1,1} \right)^2 = \left( \mu_{1,-1} \right)^2, \label{signambi}
\end{equation}
where the matrix element $\mu_{k,k'}$ is defined by
\begin{equation}
 \mu_{k,k'} \Phi^1 \equiv A^k_{1/N} A^{k'}_{-1/N} \Phi^{-1} \label{mu11def}.
\end{equation}
Using Eq.(\ref{opepara1})-Eq.(\ref{opepara2}), one finds:  
\begin{equation}
 \mu_{1,1} = \lambda^{1,1}_2 h_{1},\quad  \mu_{1,-1} = \frac{2}{N},
\label{resmu}
\end{equation}
where the structure constant $\lambda^{1,1}_2$ is given by
Eq.~(\ref{struct}), and the zero mode eigenvalue $h_1$, 
defined in Eq.~(\ref{zeroeigen}), is fixed to:
\begin{equation}
h_1^2=\frac{2}{N(N-1)}.
\label{zeroeigenfixed}
\end{equation} 
Once  the conformal dimension and the zero eigenvalue Eq.(\ref{zeroeigenfixed})
have been determined, the operator $\chi^{0}_{-\frac{1}{N}}= A^{1}_{-\frac{1}{N}}\Phi^{-1}-A^{-1}_{-\frac
 {1}{N}}\Phi^{1}$ is a primary operator and it is put to zero,
$\chi^{0}_{-\frac{1}{N}}=0$, to make the representation irreducible. After
imposing the degeneracy at level $1/N$,  only one singlet state remains:
\begin{equation}
\varepsilon^{(N)}=A^{1}_{-\frac{1}{N}}\Phi^{-1}=A^{1}_{-\frac{1}{N}}\Phi^{-1},
\end{equation}  
with dimension
\begin{equation}
\Delta_{\varepsilon^{(N)}}=\frac{2}{N+2}.
\end{equation} 

\subsection{Independent states at the second level of the $\varepsilon^{(N)}$ operator.}
Using systematically the commutation relations
Eq.(\ref{psi_2})-Eq.(\ref{psi_vir}), one can show that at the second level
of the $\varepsilon^{(N)}$ operator there are only two independent states.
All the operators at the second level can indeed be obtained by a linear
combination of the following two states:
\begin{equation}
\left(A^{1}_{-\frac{1}{N}-1}A^{-1}_{\frac{1}{N}-1}+A^{-1}_{-\frac{1}{N}-1}A^{1}_{\frac{1}{N}-1}\right)\varepsilon^{(N)}; 
\quad \mbox{and} \quad \left(A^{1}_{-\frac{1}{N}-2}A^{-1}_{\frac{1}{N}}+A^{-1}_{-\frac{1}{N}-2}A^{1}_{\frac{1}{N}}\right)\varepsilon^{(N)}.
\end{equation}
The action of the above parafermionic modes in the $\Phi^{\pm 1}$ module is
shown in Fig.(\ref{phi1modulo}).

Now, in order to make the connection with $SLE$, we express the Virasoro
operators $L_{-1}^{2}$ and $L_{-2}$ in terms of combination of
parafermionic modes.
Using the results Eq.(\ref{resmu})and Eq.(\ref{zeroeigen}) in the relations
(\ref{psi_2})-(\ref{psi_vir}), we can express the $L_{-1} \varepsilon$ operator
as:
\begin{equation}
L_{-1}\varepsilon^{(N)}=\frac{4}{N+2}\left(A^{-1}_{-1/N-1} \Phi^{1}+A^{1}_{-1/N-1}
  \Phi^{-1}\right).
\label{lmu}
\end{equation} 
The above relation, expressed in terms of the coset $SU(2)_k/U(1)$,
corresponds to the well known Knizhnik-Zamolodchikov equations.  Starting from
Eq.(\ref{lmu}) and keeping into account Eq.(\ref{resmu}) and
Eq.(\ref{zeroeigen}) and other similar relations coming from the structure of
degeneracies of the $\Phi^{\pm 1}$ module, the commutation relations
(\ref{psi_2})-(\ref{psi_vir}) give the following relations:
\begin{eqnarray}
L_{-1}^2 \varepsilon^{(N)}&=& 
\frac{2 N}{(N+2)^2}\left(A^{-1}_{-1/N-1}A^{1}_{1/N-1}+
  A^{-1}_{-1/N-1}A^{1}_{1/N-1}\right)\varepsilon^{(N)} +\frac{4}{N}\left(A^{1}_{1/N-2} \Phi^{-1}+
  A^{-1}_{-1/N-2}\Phi^{1}\right)
\label{lmu2} \\
L_{-2}\varepsilon^{(N)}&=&\frac{N}{N+2}\left(A^{-1}_{-1/N-1}A^{1}_{1/N-1}+
  A^{-1}_{-1/N-1}A^{1}_{1/N-1}\right)\varepsilon^{(N)}+\frac{4(N+1)}{N(N+2)}\left(A^{1}_{1/N-2} \Phi^{-1}+
  A^{-1}_{-1/N-2}\Phi^{1}\right).
\label{l2}
\end{eqnarray}
From the above relation, the Eq.(\ref{para_nul_vect_1}) is easily obtained.
Note that in the case of $N=3$ (the module $\Phi^{\pm 1}$ is not present in
the $N=2$ theory), the parafermionic algebra provides an additional relation
between states at the second level of $\varepsilon^{(3)}$. Indeed using the relations
$\{\Psi^{1}\Psi^{1}\}\Phi^{-1}$ and the fact that $\Psi^{1}\Psi^{1}\to
\Psi^{-1}$, valid for $N=3$, one obtains:
\begin{equation}
A^{1}_{-4/3}L_{-1}\Phi^{-1}=\frac{1}{5}A^{-1}_{-7/3}\Phi^{1}-\frac{1}{15}A^{1}_{-7/3}\Phi^{-1}
\label{z3p}
\end{equation}
Using Eq.(\ref{z3p}) and Eq.(\ref{lmu}) in Eq.(\ref{l2})and in
Eq.(\ref{z3p}), one gets:
\begin{equation}
\left(L_{-2}-\frac{5}{6}L_{-1}^2\right) \varepsilon^{(N=3)}=0
\end{equation}
which is the relation (\ref{vir_nul_vect_1}) obtained by the study of the
minimal model $M_5$ (remember that $\varepsilon^{(N=3)}$ corresponds to
$\phi_{1,2}$ of the minimal model Kac table). 

\subsection{Commutation relations of parafermionic current modes:$R-$sector}
\label{rsector_kac}
As mentioned in the Section (\ref{Rsector}), the reader can refer to \cite{Zamo2} and
\cite{Raoul1} for a complete discussion about the construction of the disorder sector
modules and of the derivation of the commutation relations $\{\Psi^{k},
\Psi^{k'}\}R_a$. In the calculations of degeneracy we have used two types of
commutation relations: the first one is between the modes of two $\Psi^1$
chiral fields and the second one is between the $\Psi^1$ and $\Psi^k$ chiral
fields, with $k=2,3,\ldots,\frac{N-1}{2}$. Using the expansions
Eq.(\ref{rexpansion}) and Eq.(\ref{rmodes}), the  
$\{\Psi^{k},\Psi^{1}\}$ relations have the following form:
\begin{eqnarray}
  \sum_{l} D^{l}_{1-2 k/N}&&\left[A^{k}_{(n-l)/2}A^{1}_{(m+l)/2}+
    A^{1}_{(m-l)/2}A^{k}_{(n+l)/2} \right]R_{a}= \nonumber \\
&&\left[2^{1-4
      k/N}\lambda^{k,1}_{k+1}A^{k+1}_{(n+m)/2}\delta_{a,a'}+2^{-3+4
      k/N}\lambda^{k,-1}_{k-1}(-1)^{m} A^{k+1}_{(n+m)/2 }{\sf U}\right]R_{a'}
\label{rcr1}
\end{eqnarray}
where the coefficients $D^{l}_{\alpha}$ are defined from the expansion:
\begin{equation}
(1-x)^{-\alpha}(1+x)^{\alpha}=\sum_{l} D^{l}_{\alpha} x^{l},
\end{equation}
and  the matrix  ${\sf U}^k R_{a} =R_{a-k}$ changes the $R$ indices. 
For the $\{\Psi^{1},\Psi^{1}\}$ relations, which from the
Eq.(\ref{rexpansion}) establish the connection with the Virasoro generators,
one gets:
\begin{eqnarray}
  \sum_{l} D^{l}_{-1-2/N}&&\left[A^{1}_{(n-l)/2}A^{1}_{(m+l)/2}+
    A^{1}_{(m-l)/2}A^{1}_{(n+l)/2} \right]R_{a}= \nonumber \\
&&2^{1+4
      k/N}(-1)^m\left[\frac{N+2}{N}L_{(n+m)/2}+
      \kappa(n)\delta_{n+m,0}\right]{\sf U}R_{a'} 
\label{rcr2}
\end{eqnarray}
where 
\begin{equation}
\kappa(n)=\frac{n^2}{8}-\frac{(N-2)}{16 N}.
\end{equation}
Finally we will use the commutations $\{T,\Psi\}R_a$:
\begin{equation}
[L_n,A^{1}_{m/2}]=\left[(n+1)\Delta_{1}-(n-\frac{m}{2}+\Delta)\right]
A^{1}_{n+m/2}.
\label{vir_psi_r}
\end{equation}

The level structure of the modules of disorder operators is relatively simple.
There are only integer and half-integer levels, and there exists zero modes
for all the operators $\{ \Psi^{q} \}$ acting on the $N-$uplet of disorder
operators.  From the expansion (\ref{rmodes}) there are $\lfloor N/2 \rfloor$
zero modes $A^{k}_{0}$ (with $k=1,2,\ldots,\frac{N-1}{2}$), associated with
the parafermion $\Psi^k$ which acts between the $N$ summit of the module:
\begin{equation}
 A^{k}_{0}R_{a} = h_{k} \, {\sf U}^{2k}R_{a}. \label{Rzero1} 
\end{equation}

\begin{figure}
\begin{center}
 \leavevmode
 \includegraphics[scale=0.5]{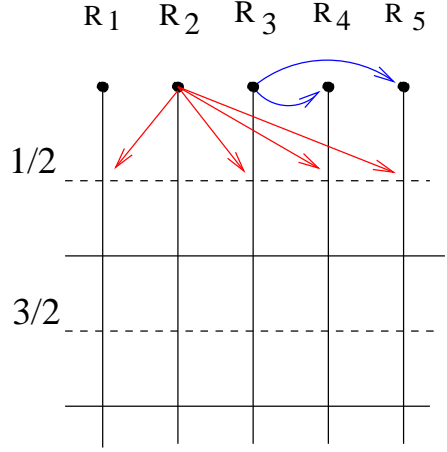}
 \end{center}
 \protect\caption[3]{\label{rmodulo} Representation module of $R_a$ fields for
   $N=5$. The arrows show the action of the parafermionic modes. The zero mode
 action between the summit of the module are illustrated by the blue arrows.}
\end{figure}

This defines the eigenvalues $h_{k}$. We recall that ${\sf U} R_{a} =
R_{a-1}$.  The eigenvalues $h_{k}$ characterize, together with the conformal
dimension, each representation $R_a$. Like the case of the $Z_N$ sector, the
formula for the conformal dimension (\ref{disorder_kac}) together with the
zero modes $h_k$ are easily derived by solving the system of equations
obtained by setting $n=m=0$ in the Eq.(\ref{rcr1}) and Eq.(\ref{rcr2}).
Here we focus our attention on the module of the operator $R_a^{(0)}$ with
dimension:
\begin{equation}
\Delta^{R}_{0}=\frac{N^2+N-2}{16(N+2)}.
\label{dimro}
\end{equation}

\subsection{Independent states at the second level of the $R^{(0)}$ operator.}
We have analyzed more in detail the structure of the module of the $R^{(0)}$
disorder operator. Putting the values $n=0, m=-1$ in the   Eq.(\ref{rcr1}) and Eq.(\ref{rcr2}), and using
Eq.(\ref{dimro}) together with the value of the zero mode $y_1$  
\begin{equation}
y_1=2^{-2(N-1)/N} \sqrt{N} 
\end{equation}
caractherizing the module of $R^{(0)}_a$, it can be shown that
\begin{equation}
A^{k}_{-1/2}R^{(0)}_a=0 \quad k=1,2,..\lfloor N/2 \rfloor
\label{degr}
\end{equation}
for each $k$. This simplifies greatly the
computation of the relations at the second level of this operator, in which
 we are interested.

 At the first level, by setting $n=-1, m=-1$ in the Eq.(\ref{rcr2}) and using
 Eq.(\ref{degr}), one gets:
\begin{equation}
L_{-1}R^{(0)}_{a}=2^{-4-2/N}\sqrt{N} A^{1}_{-1} R^{(0)}_{a+1}
\label{1levelr}
\end{equation}
Combining the above relation and the Eq.(\ref{vir_psi_r}) with $n=-1$ and
$m=-2$, we find:
\begin{equation}
L_{-1}^2 R^{(0)}_{a}=2^{-1-2/N}\frac{N+1}{\sqrt{N}}A^{1}_{-2}R^{(0)}_{a+1}-
2^{-2-4/N}N A^{1}_{-1}A^{-1}_{-1}R^{(0)}_{a}.
\label{lm12r}
\end{equation}
Finally, to express the operator $L_{-2}$ we use Eq.(\ref{rcr2}) with $n=-2$
and $m=-2$, obtaining:
\begin{equation}
L_{-2}
R^{(0)}_{a}=2^{-1-2/N}\frac{N+2}{\sqrt{N}}A^{1}_{-2}R^{(0)}_{a+1}-2^{-4/N}
\frac{N}{N+2} A^{1}_{-1}A^{-1}_{-1}R^{(0)}_{a}.
\label{lm2r}
\end{equation}
The Eq.(\ref{para_nul_vect_2}) is then directly derived from the above
relations.

As in the case of the $\varepsilon^{(N=3)}$ operator, we have verified that,
in this approach, we find the known results for the $\phi_{2,1}$ operator of
the $M_5$ model (remember that for $N=3$, $R^{(0)}=\phi_{2,1}$). Indeed, using
the fact that $\Psi^{2}=\Psi^{-1}$ for $N=3$, we can extract from the
Eq.(\ref{rcr1}) with $n=-2,m=-2$ the following relation:
\begin{equation}
 A^{1}_{-1}A^{-1}_{-1}R^{(0)}_{a}=2^{2/3}\frac{\sqrt{3}}{3} A^{1}_{-2}R^{(0)}_a.
\label{reln3}
\end{equation}  
Using Eq.(\ref{reln3}) in Eq.(\ref{lm12r}) and in Eq.(\ref{lm2r}), we obtain:
\begin{equation}
\left(L_{-2}-\frac{6}{5}L_{-1}^2\right) R^{(0)}_a=0
\end{equation}
in agreement with Eq.(\ref{vir_nul_vect_2}).

\section{Boundary states.}
\label{boundarystates}

One of the main results of the Boundary conformal field theory
\cite{Cardy_bcft1,Cardy_bcft2} is the
bijection between the possible conformally invariant boundary conditions and
the bulk primary operators.  In particular, the allowed boundary states are
expressed as linear combinations of bulk primary operators. The coefficients
of such expansion are directly related to the entries of modular
transformation matrix $\mathcal{S}$.
Reminiscent of the coset construction of the parafermionic theories,
$Z_N=SU(2)_N/U(1)$, it is  convenient to classify the bulk operators with the
notation  
$|j,m>$, with $j=0,1,\cdots$
and $m=-\lfloor N/2 \rfloor,\lfloor N/2 \rfloor+1,\cdots, \lfloor N/2
\rfloor$. The boundary states $\overline{|l,m>}$ are then defined by the
  formula:
\begin{equation}
\overline{|l',m'>}=\sum_{l,m}\frac{\mathcal{S}^{l',m'}_{l,m}}{\mathcal{S}^{0,0}_{l,m}}|l,m>
\label{cardy}
\end{equation}

The three states Potts model represents a typical example where these
results apply. In order to fix the notation and to facilitate the comparison  with the
other $Z_N$ theories, we reproduce below the construction of the boundary states
for this model. We use the same conventions as in \cite{Cardy_bcft2}.
 
The table of the primary fields of the $Z_3$ theory with the
associated characters is:
\begin{center}
\begin{tabular}{c|c|c}
 Field \hspace{0.5cm} & $\Delta$ \hspace{0.5cm} & \hspace{0.2cm} $|l, m>$ \hspace{0.5cm} \\
\hline & & \\ 
$\Phi^{0}$\hspace{0.5cm}& 0 \hspace{0.5cm}& \hspace{0.2cm} $|0,0>$ \hspace{0.5cm} \\
& & \\
$\Psi^{1}$\hspace{0.5cm}& 2/3 \hspace{0.5cm}& \hspace{0.2cm} $|0,1>$ \hspace{0.5cm} \\
& & \\
$\Psi^{-1}$\hspace{0.5cm}& 2/3 \hspace{0.5cm}& \hspace{0.2cm} $|0,-1>$ \hspace{0.5cm} \\
& & \\
$\Phi^{1} $\hspace{0.5cm}& 1/15 \hspace{0.5cm}& \hspace{0.2cm} $|1,1>$ \hspace{0.5cm} \\
& & \\
$\varepsilon^{(N=3)}$\hspace{0.5cm}& 2/5 \hspace{0.5cm}& \hspace{0.2cm} $|1,0>$ \hspace{0.5cm} \\
& & \\
$\Phi^{-1} $\hspace{0.5cm}& 1/15 \hspace{0.5cm}& \hspace{0.2cm} $|1,-1>$ \hspace{0.5cm}
\end{tabular}
\end{center}
In the basis
$(|0,0>,|1,0>,|0,1>,|1,1>,|0,-1>,|1,-1>)$
the modular transformation matrix $\mathcal{S}\equiv
\mathcal{S}^{l',m'}_{l,m}$ takes the form:
\begin{equation}
\mathcal{S}^{m'}_{m}=\frac{1}{\sqrt{3}}\left( \begin{array}{c c c}
                              s^{l'}_{l}&  s^{l'}_{l}&
                              s^{l'}_{l} \\
s^{l'}_{l}& \omega s^{l'}_{l} & \omega^2
s^{l'}_{l} \\ 
s^{l'}_{l}& \omega^2 s^{l'}_{l} & \omega s^{l'}_{l}
\end{array}
\right)
\end{equation}
where $\omega=\exp{i \pi/3}$ and 
\begin{equation}
s^{l'}_{l}=\frac{1}{\sqrt{5}}\left( \begin{array}{c c} 
                                    \sin{\pi/5} & \sin{2\pi/5} \\
                                   \sin{2\pi/5} & -\sin{\pi/5}
                                   \end{array} \right) 
\end{equation}
In order to identify their lattice realization, it is
important to study the behavior of the boundary states under a  $Z_3$
rotation. Using the formula (\ref{cardy}), one can observe that under a $Z_3$
rotation the boundary states transform as:
\begin{eqnarray}
&&\overline{|0,0>}\to \overline{|0,1>}\to \overline{|0,-1>}\to
\overline{|0,0>} \nonumber \\
&&\overline{|1,0>}\to \overline{|1,1>}\to \overline{|1,-1>}\to \overline{|1,0>}
\end{eqnarray}
Identifying the boundary state $\overline{|0,0>}$ with the boundary conditions in
which all the spin on the real axis take the value $A$, the above
transformations suggests the identification of the state $\overline{|0,1>}$
($\overline{|0,-1>}$) to the situation in which all the spins on the real axis
take the value $B$ ($C$). In consistence with the above transformations, the
state $\overline{|1,0>}$ has been shown to correspond to the state in which
the spins can take with equal probability the values $B$ or $C$. 
From this identification one concludes that the b.c.c.  operator
$\psi_{\varepsilon}$ generates the boundary conditions $A|B+C$ discussed
before.

The results shown here for $Z_3$ generalize directly to the $Z_N$ theories.
The modular transformation properties of these theories have been analyzed in \cite{Gepner}.  For sake of simplicity we consider here the case $Z_5$.  In
general, the number of irreducible representation of the Virsasoro algebra is
infinite for $N>3$. However, there is a system of principal Virasoro primary fields
which appear at the first levels in each module of $\Phi^{q}$. The table of the
principal primary fields of the $Z_5$ theory is the following:
\begin{center}
\begin{tabular}{c|c|c}
 Field \hspace{0.5cm} & $\Delta$ \hspace{0.5cm} & \hspace{0.2cm} $|l, m>$ \hspace{0.5cm} \\
\hline & & \\ 
$\Phi^{0}$\hspace{0.5cm}& 0 \hspace{0.5cm}& \hspace{0.2cm} $|0,0>$ \hspace{0.5cm} \\
& & \\
$\Psi^{1}$\hspace{0.5cm}& 4/5 \hspace{0.5cm}& \hspace{0.2cm} $|0,1>$ \hspace{0.5cm} \\
& & \\
$\Psi^{-1}$\hspace{0.5cm}& 4/5 \hspace{0.5cm}& \hspace{0.2cm} $|0,-1>$ \hspace{0.5cm} \\
& & \\
$\Psi^{2}$\hspace{0.5cm}& 6/5 \hspace{0.5cm}& \hspace{0.2cm} $|0,2>$ \hspace{0.5cm} \\
& & \\
$\Psi^{-2}$\hspace{0.5cm}& 6/5 \hspace{0.5cm}& \hspace{0.2cm} $|0,-2>$ \hspace{0.5cm} \\
& & \\
$\Phi^{1} $\hspace{0.5cm}& 3/35 \hspace{0.5cm}& \hspace{0.2cm} $|1,1>$ \hspace{0.5cm} \\
& & \\
$\varepsilon^{(N=5)}$\hspace{0.5cm}& 2/7 \hspace{0.5cm}& \hspace{0.2cm} $|1,0>$ \hspace{0.5cm} \\
& & \\
$\Phi^{-1} $\hspace{0.5cm}& 3/35 \hspace{0.5cm}& \hspace{0.2cm} $|1,-1>$ \hspace{0.5cm}\\
& & \\
$A^{1}_{-2/5}\Phi^{1} $\hspace{0.5cm}& 17/35 \hspace{0.5cm}& \hspace{0.2cm}$|1,2>$ \hspace{0.5cm} \\
& & \\
$A^{-1}_{-2/5}\Phi^{-1} $\hspace{0.5cm}& 17/35 \hspace{0.5cm}& \hspace{0.2cm}$|1,-2>$ \\
& & \\
$\Phi^{2} $\hspace{0.5cm}& 2/35 \hspace{0.5cm}& \hspace{0.2cm} $|2,2>$ \hspace{0.5cm} \\
& & \\
$A^{-1}_{-3/5}\Phi^{2} $\hspace{0.5cm}& 23/35 \hspace{0.5cm}& \hspace{0.2cm} $|2,1>$\hspace{0.5cm} \\
& & \\
$A^{-2}_{-4/5}\Phi^{2} $\hspace{0.5cm}& 6/7 \hspace{0.5cm}& \hspace{0.2cm}$|2,0>$\hspace{0.5cm} \\
& & \\
$A^{1}_{-3/5}\Phi^{-2} $\hspace{0.5cm}& 23/35 \hspace{0.5cm}& \hspace{0.2cm}$|2,-1>$\hspace{0.5cm} \\
& & \\
$\Phi^{2} $\hspace{0.5cm}& 2/35 \hspace{0.5cm}& \hspace{0.2cm}$|2,-2>$\hspace{0.5cm} 
\end{tabular}
\end{center}
In the basis
$(|0,0>,|1,0>,|2,0>,|0,1>,|1,1>,|2,1>,|0,2>,|1,2>,|2,2>,|0,-1>,|1,-1>,|2,-1>,|0,-2>,|1,-2>,|2,-2>)$,
the modular transformation matrix $\mathcal{S}\equiv
\mathcal{S}^{l',m'}_{l,m}$ takes the form:
\begin{equation}
\mathcal{S}^{m'}_{m}=\frac{1}{\sqrt{5}}\left( \begin{array}{c c c c c}
                              s^{l'}_{l}&  s^{l'}_{l}&
                              s^{l'}_{l} & s^{l'}_{l}& s^{l'}_{l}\\
s^{l'}_{l}& \omega s^{l'}_{l} & \omega^2
s^{l'}_{l} & \omega^{-1} s^{l'}_{l} & \omega^{-2} s^{l'}_{l} \\ 
s^{l'}_{l}& \omega^2 s^{l'}_{l} &- \omega^{4} s^{l'}_{l} & \omega^{-2} s^{l'}_{l} &
\omega^{-4} s^{l'}_{l}\\
s^{l'}_{l}& \omega^{-1} s^{l'}_{l} & \omega^{-2} s^{l'}_{l} & \omega s^{l'}_{l} &
\omega^{2} s^{l'}_{l}\\
s^{l'}_{l}& \omega^{-2} s^{l'}_{l} & \omega^{-4} s^{l'}_{l} & \omega^{2} s^{l'}_{l} &
\omega^{4} s^{l'}_{l}
\end{array}
\right)
\end{equation}
where $\omega=\exp{i \pi/5}$ and 
\begin{equation}
s^{l'}_{l}=\frac{1}{\sqrt{7}}\left( \begin{array}{c c c} 
                                    \sin{\pi/7} & \sin{2\pi/7} &\sin{2\pi/7} \\
                                   \sin{2\pi/7} & \sin{4\pi/7}& \sin{6\pi/7}\\
                                   \sin{3\pi/7} & \sin{6\pi/7}& \sin{4\pi/7}
                                   \end{array} \right) 
\end{equation}
The corresponding boundary states
$\overline{|0,0>}$ and $\overline{|1,0>}$ transforms, under a $Z_5$
transformation, in the following way:
\begin{eqnarray}
&&\overline{|0,0>}\to \overline{|0,1>}\to \overline{|0,2>}\to
\overline{|0,-2>}\to  \overline{|0,-1>}\nonumber \\
&&\overline{|1,0>}\to \overline{|1,1>}\to \overline{|1,2>}\to \overline{|1,-2>}\to \overline{|1,-1>}.
\end{eqnarray}
It is clear that the case $Z_5$, and $Z_N$ in general, represents,  from the point of view
 of the properties of the boundary states, a direct generalization of the case
 $Z_3$. It is then natural to identify the boundary state $\overline{|1,0>}$
 with the state in which the spins can take with equal probability the values
 $B,C,D,E$. Consequently, the b.c.c operator $\psi_{\varepsilon}$ is expected
 to generates the boundary conditions $A|B+C+D+E$ discussed in section
 \ref{lattice model}.

%\end{spacing}

\begin{thebibliography}{10}
\bibitem{Smirnov} S.~Smirnov,
 \newblock C.R. Acad. Sci. Paris {\bf 333}, 239 (2001).

\bibitem{Lawler} G.~Lawler, O. Schramm and W. Werner,
 \newblock Ann. Prob.{\bf 32}, 939 (2004).

\bibitem{Schramm} O.~Schramm and S.~Sheffield,
 \newblock arXiv:math.Pr/0605237.

\bibitem{Bernard_review} M.~Bauer and D.~Bernard, \newblock Phys. Rept. {\bf 432}, 115 (2006).

\bibitem{Bernard_connection1} M.~Bauer and D.~Bernard, \newblock Comm. Math.
  Phys.{\bf 239}, 493 (2003).

\bibitem{Bernard_connection2} M.~Bauer and D.~Bernard, \newblock
  Phys. Lett. B{\bf 543}, 135 (2002).

\bibitem{Bernard_connection3} M.~Bauer and D.~Bernard, \newblock
  Phys. Lett. B{\bf 557}, 309 (2003).

\bibitem{Bernard_connection4} M.~Bauer and D.~Bernard, \newblock
  Phys. Lett. B{\bf 583}, 324 (2004).

\bibitem{Affleck} I.~Affleck and F.~D.~M.~Haldane
 \newblock Phys. Rev. B. {\bf 36}, 5291 (1987).


\bibitem{Zamo1}  V.~A.~Fateev  and A.~B.~Zamolodchikov,\newblock
                      Sov.~Phys.~JETP {\bf 62}, 215 (1985).

\bibitem{Zama_lat2} V.~A.~Fateev and A.~B.~Zamolodchikov, \newblock 
 Phys.~Lett.~JETP {\bf 92A}, 37 (1982).


\bibitem{Cardy_bcft1} J.~Cardy, \newblock 
Nucl.\ Phys.\ B {\bf 240}, 514 (1984).

\bibitem{Cardy_bcft2} J.~Cardy, \newblock 
Nucl.\ Phys.\ B {\bf 324}, 581 (1989).

\bibitem{Rasmussen1} J. ~Rasmussen
 \newblock Lett. Math. Phys.  {\bf 68}, 41-52 (2004).
 
 \bibitem{Rasmussen2} J.~Nagi, J.~ Rasmussen
 \newblock Nucl. Phys. B {\bf 704}, 475-489 (2005).
 
 \bibitem{Rasmussen3} J.~Rasmussen
 \newblock hep-th/0409026.


\bibitem{Ludwig} E.~Bettelheim, I.~A.~Gruzberg,A.~W.~W.~Ludwig and P.~Wiegmann
 \newblock Phys. Rev. Lett. {\bf 95}, 170602 (2005).

\bibitem{Walter} Q.~Kager and B.~ Nienhuis,
 \newblock J. Stat. Phys.{\bf 115}, 1149 (2004).

\bibitem{Cardy_review} J.~Cardy,
 \newblock Annals Phys.{\bf 318}, 81 (2005).


\bibitem{Schramm_2} O.~Schramm, \newblock Israel. J.Math. {\bf 118}, 221 (2000).

\bibitem{Friedan} D.~Friedan, Z.~ Qiu and S.~Shenker.
 \newblock Phys. Rev. Lett. {\bf 52}, 1575 (1984).


\bibitem{Zamo2}  V.~A.~Fateev  and A.~B.~Zamolodchikov,\newblock
        
              Sov.~Phys.~JETP {\bf 63}, 913 (1986).

\bibitem{Raoul1} Vl.~Dotsenko, J.L. Jacobsen, and R.~Santachiara, \newblock 
Nucl.\ Phys.\ B {\bf 656}, 259 (2003).

\bibitem{Raoul2} Vl.~Dotsenko, J.L. Jacobsen, and R.~Santachiara, \newblock 
Nucl.\ Phys.\ B {\bf 664}, 477 (2003).

\bibitem{Raoul3} Vl.~Dotsenko, J.L. Jacobsen, and R.~Santachiara, \newblock 
Phys.\ Lett.\ B {\bf 584}, 186 (2004).

\bibitem{Raoul4} Vl.~Dotsenko, J.L. Jacobsen, and R.~Santachiara, \newblock 
Nucl.\ Phys.\ B {\bf 679}, 464 (2004).

\bibitem{Zama_lat} A.~B.~Zamolodchikov, \newblock 
 Sov.~Phys.~JETP {\bf 48}, 168 (1978).

\bibitem{Dotsi_lat} V.~S.~Dotsenko, \newblock 
 Sov.~Phys.~JETP {\bf 48}, 546 (1978).

\bibitem{Fradkin_lat} E.~Fradkin and J.~Kadanoff \newblock 
 Nucl. Phys. B {\bf 170}, FS1.1. (1980).


\bibitem{Gepner} D.~Gepner, Z.~Qiu \newblock 
Nucl.\ Phys.\ B {\bf 285}, 423 (1987).


\bibitem{Riva} V~Riva, J.~Cardy, \newblock 
J. Stat\ Mech. {\bf 0612}, P001 (2006).


\bibitem{Cardy_potts} A.~Gamsa, J.~Cardy, \newblock 
cond-mat/0705.1510

\bibitem{Bauer_R} H.~Saleur, M.~Bauer \newblock 
Nucl.\ Phys.\ B {\bf 320}, 591 (1989).




\end{thebibliography}
\end{document}